\documentclass[print]{aa} 
\usepackage{savesym}%
\usepackage{graphicx}%
\usepackage{lineno}%
\usepackage{ulem}
\usepackage{mathtools}%
\usepackage{footnote}
\usepackage{amsmath}%
\usepackage{xspace}%
\usepackage{units}%
\usepackage{relsize}%
\usepackage[dvipsnames]{xcolor}
\usepackage{tikz}%
\usepackage{amssymb,amsfonts,amsmath,amstext,amsgen,amsopn,amsxtra,indentfirst,times}%
\usetikzlibrary{positioning}%
\usetikzlibrary{shapes,arrows}%
\usepackage{nicefrac}%
\usepackage{array}
\usepackage{multirow}
\usepackage{natbib}

\DeclareFontFamily{U}{mathb}{\hyphenchar\font45}
\DeclareFontShape{U}{mathb}{m}{n}{
      <5> <6> <7> <8> <9> <10> gen * mathb
      <10.95> mathb10 <12> <14.4> <17.28> <20.74> <24.88> mathb12
      }{}
\DeclareSymbolFont{mathb}{U}{mathb}{m}{n}
\DeclareMathSymbol{\Earth}{3}{mathb}{"43}

\usepackage{tabularx}  
\usepackage{ragged2e}  
\newcolumntype{Y}{>{\RaggedRight\arraybackslash}X} 
\usepackage{booktabs}  

\def\widesplit#1{%
\cleardoublepage
\def\row##1##2{##1}%
#1%
\\

\def\row##1##2{##2}%
#1%
\clearpage
}

\newcommand{\RE}{R$_{\rm \Earth}$\xspace}
\newcommand{\ME}{M$_{\rm \Earth}$\xspace}
\newcommand{\CE}{c$_{\rm E}$\xspace}
\newcommand{\pg}{${\rm d}p/{\rm d}z$\xspace}
\newcommand{\rc}{$r_{\rm core}$\xspace}
\newcommand{\rcM}{r_{\rm core}\xspace} 
\newcommand{\fesisun}{{\rm Fe}/{\rm Si}_{\rm sun}\xspace}
\newcommand{\mgsisun}{{\rm Mg}/{\rm Si}_{\rm sun}\xspace}
\newcommand{\fesima}{{\rm Fe}/{\rm Si}_{\rm mantle}\xspace}
\newcommand{\mgsima}{{\rm Mg}/{\rm Si}_{\rm mantle}\xspace}
\newcommand{\fesi}{{\rm Fe}/{\rm Si}_{\rm bulk}\xspace}
\newcommand{\mgsi}{{\rm Mg}/{\rm Si}_{\rm bulk}\xspace}

\definecolor{Lcolor}{rgb}{0.1,0.5,0.3}

\newcommand{\rev}[1]{{\color{black} {#1}}} 
\begin{document}

\title{Outgassing on stagnant-lid super-Earths}

\author{C. Dorn\inst{1}, L. Noack\inst{2,3}, A. B. Rozel\inst{4}}
\titlerunning{Outgassing on stagnant-lid exoplanets}
\authorrunning{Dorn et al.}
          \institute{Institute of Computational Sciences, University of Zurich, Winterthurerstrasse 109, CH-8057, Zurich, Switzerland\\
              \email{cdorn@physik.uzh.ch}
              \and{Department of Reference Systems and Geodynamics, Royal Observatory of Belgium, Avenue Circulaire 3, 1180 Brussels, Belgium}
              \and{Institute of Geological Sciences, Free University Berlin, Malteserstr. 74-100, 12249 Berlin, Germany}
              \and{Institute of Geophysics, Department of Earth Sciences, ETH Zurich,
  Sonneggstrasse 5, 8092 Zurich, Switzerland}
             }


\abstract{}
   {We explore volcanic outgassing on purely rocky, stagnant-lid exoplanets of different interior structures, compositions and thermal states. We focus on planets in the mass range of 1--8 \ME (Earth masses). We derive scaling laws to  {quantify} first- and second-order influences of these parameters on volcanic outgassing after 4.5 Gyrs of evolution.}
{Given commonly observed astrophysical data of super-Earths, we identify a range of possible interior structures and compositions by employing Bayesian inference modelling. The astrophysical data comprises mass, radius, and bulk compositional constraints, i.e. ratios of refractory element abundances are assumed to be similar to stellar ratios. The identified interiors are subsequently used as input
for two-dimensional (2-D) convection models to study partial melting, depletion, and outgassing rates of CO$_2$.}
{In total, we model depletion and outgassing for an extensive set of more than  {2300} different super-Earth cases. We find that there is a mass range for which outgassing is most efficient ($\sim$2--3 \ME, depending on thermal state) and an upper mass where outgassing becomes very  inefficient ($\sim$5--7 \ME, depending on thermal state). At small masses (below 2--3~\ME) outgassing positively correlates with planet mass, since it is controlled by mantle volume. At higher masses (above 2--3~\ME), outgassing decreases with planet mass, which is due to the increasing pressure gradient that limits melting to shallower depths. In summary, depletion and outgassing are mainly influenced by planet mass and thermal state. Interior structure and composition only moderately affect outgassing.  {The majority of outgassing occurs before 4.5 Gyrs, especially for planets below 3~\ME.} }
{ We conclude that for stagnant-lid planets,  {(1) compositional and structural properties have secondary influence on outgassing compared to planet mass and thermal state, and (2) confirm that there is} a mass range for which outgassing is most efficient and an upper mass limit, above which no significant outgassing can occur. \rev{Our predicted trend of CO$_2$-atmospheric masses can be observationally tested for exoplanets.} These findings and our provided scaling laws are an important step  {in order} to provide interpretative means for upcoming missions  {such as} 
JWST and E-ELT, that aim at characterizing exoplanet atmospheres.}
   \keywords{}

   \maketitle

\section{Introduction}

\begin{figure*}
  \centering
  \includegraphics[width=.7\hsize]{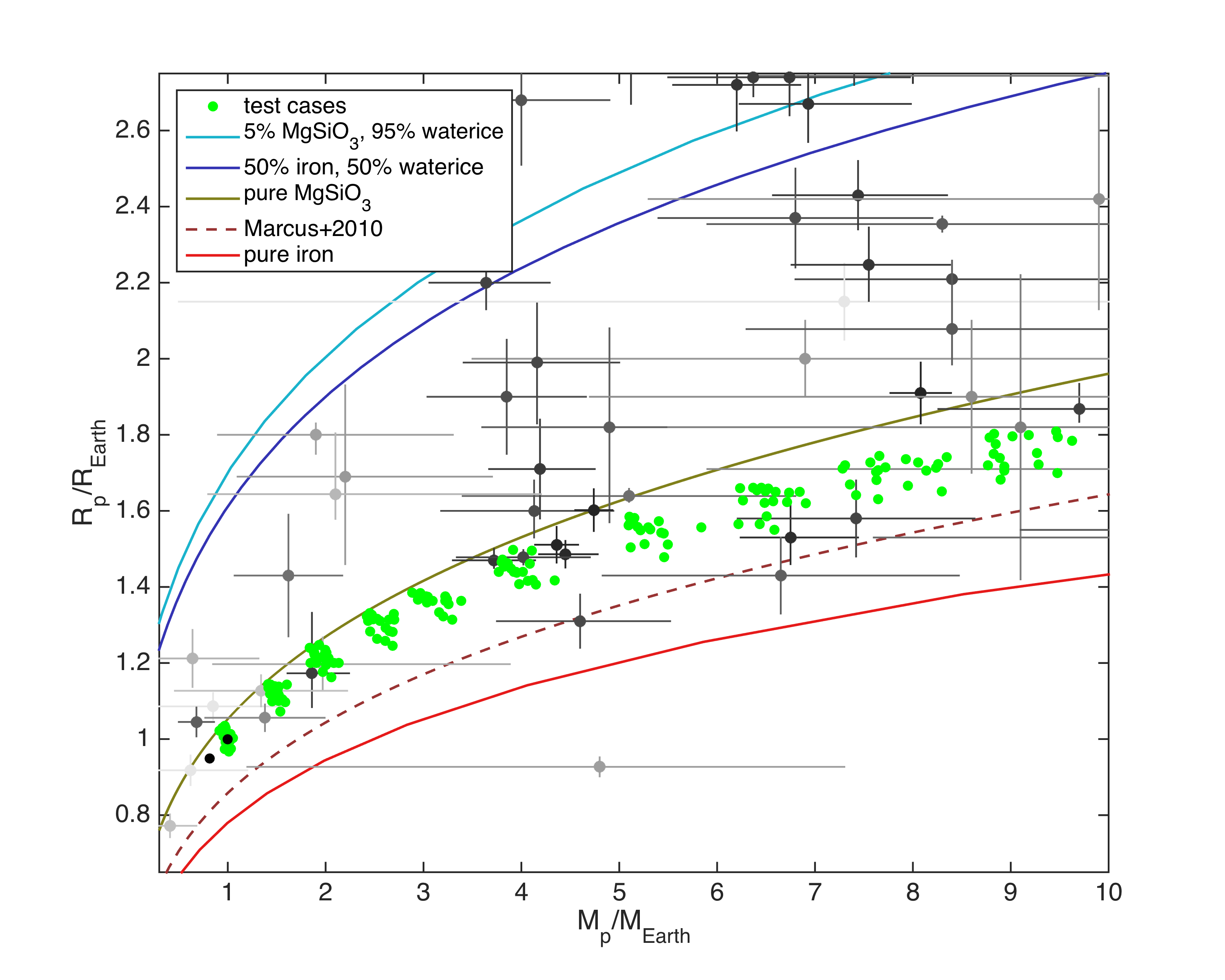}
  \caption{Mass-radius diagram for planets below 2.7 \RE and 10 \ME (180 Super-Earths shown). Transparencies of the black points scale with the relative error on planet mass. Green dots represent synthetic planets used in our study. The dashed curve denotes the minimum radius predicted for maximum mantle stripping due to giant impacts \citep{marcus2010minimum}.}\label{fig:MR}
    \end{figure*}

\sloppy
Super-Earths are among the most abundant exoplanets and are characterized by small volatile fractions \citep[e.g.,][]{dressing2015, fulton2017}. Super-Earths have planet masses and radii that exceed the diversity of the Solar System planets (Figure \ref{fig:MR}). Our knowledge of the variability of their interiors is limited, because data (e.g., mass and radius) are few and do allow for very different interior structures and compositions. The only parts of exoplanets that can be directly probed are their atmospheres. So far, there are only few small-mass planets (GJ1214b, HD97658b, 55Cnc e, GJ1132b) for which constraints on their atmospheres are available. However, near future spectroscopic observations (e.g., E-ELT, JWST) will allow us to gain detailed insights into the atmospheric compositions for a number of super-Earths. 

The anticipated diversity of atmospheres on super-Earth exoplanets is subject to planet formation and evolution processes \citep{leconte2015anticipated}. Different processes can shape the thickness and chemical make-up of an  atmosphere: gas accretion from the stellar nebular, atmospheric enrichment by the disruption of planetesimals, outgassing from an early magma ocean or long-term out- and in-gassing processes, and  hydrodynamic escape. The understanding of these processes is crucial for the interpretation of  atmospheric characteristics inferred from observations. 
Here, we focus on volcanic outgassing that can constantly release volatiles on geological timescales into the atmosphere that were once trapped in the mantle. Volcanic outgassing can be the origin of enriched atmospheres, that \citet{dorn2017submitted} identified to be likely dominating those planets of small-masses and warm to hot equilibrium temperatures. The importance of volcanic outgassing on observed super-Earths is ongoing research. We anticipate that the diversity in planetary interiors and thermal states may significantly influence volcanic activity and consequently the thicknesses of outgassed atmospheres, which we will address in this study.

\begin{figure}
  \centering
  \includegraphics[width=1\hsize, trim= .7cm 0cm 1.8cm 0cm, clip]{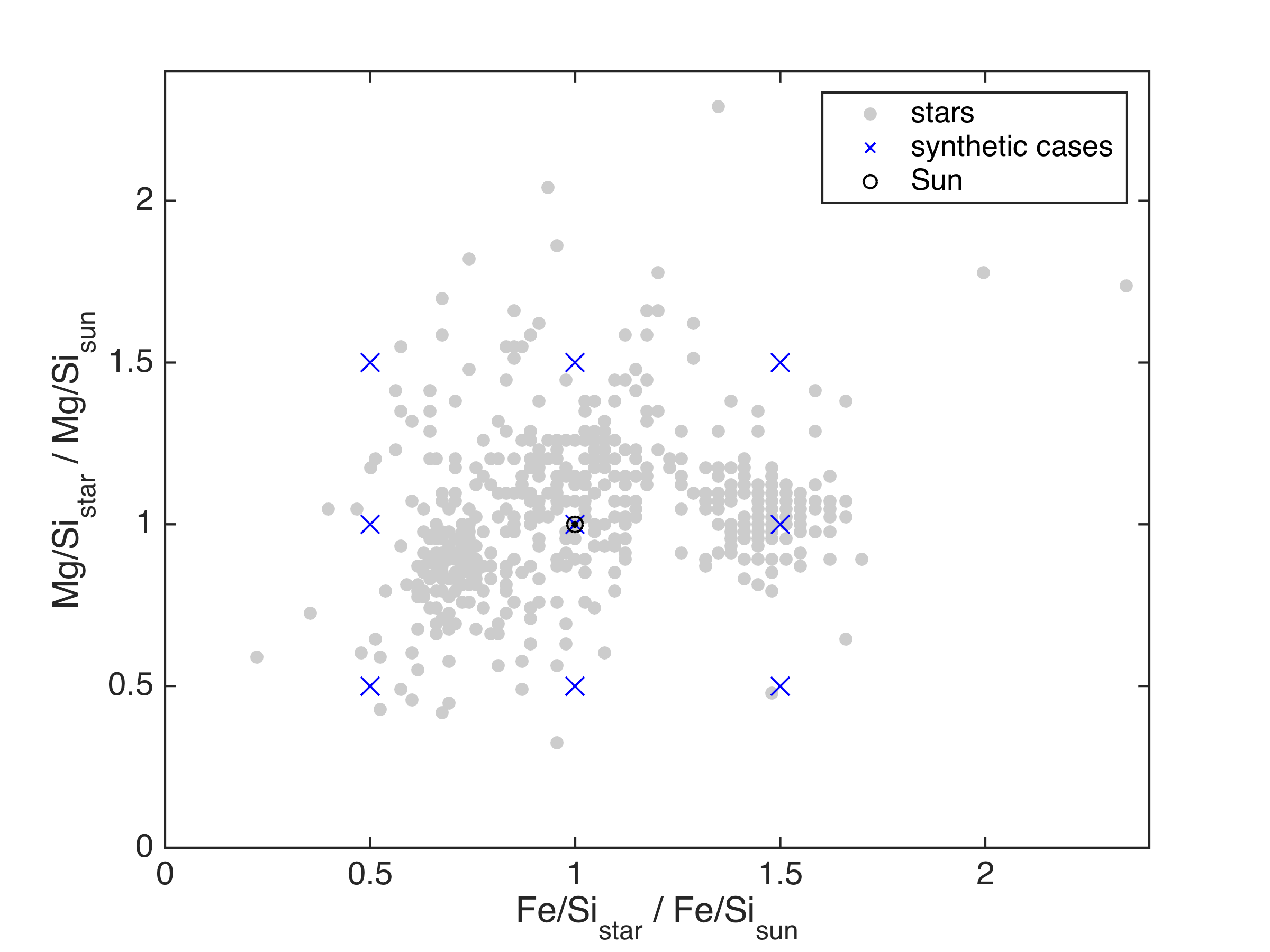}
  \caption{Stellar abundances Fe/Si$_{\rm star}$ and Mg/Si$_{\rm star}$ for stars within 150 pc based on the Hypatia catalog \citep{hinkel2014stellar}. Values are relative to solar estimates (Fe/Si$_{\rm Sun}$ = 1.69 and Mg/Si$_{\rm Sun}$= 0.89 based on \citet{lodders2003solar}). Blue crosses denote synthetic planetary bulk abundances used in our study (Fe/Si$_{\rm bulk}$ = \{0.5, 1., 1.5\} $\times$ Fe/Si$_{\rm Sun}$ and Mg/Si$_{\rm bulk}$ = \{0.5, 1., 1.5\} $\times$ Mg/Si$_{\rm Sun}$)
  }\label{fig:Ratios}
    \end{figure}

The diversity in interior structures and compositions for observed exoplanets is generally expected to be  large. For rocky exoplanets, despite  the given data of planetary masses and radii, there is significant ambiguity on possible core sizes and mantle compositions. This ambiguity can be significantly reduced by accounting for possible correlations between stellar and planetary compositions, specifically their relative abundances of rock-forming elements (e.g., Fe, Si, Mg) \citep{dorn2015can}. The observed relative abundances on Fe/Si and Mg/Si of planet-hosting stars have limited variability (Figure \ref{fig:Ratios}). Here, we assume that the variability of stellar abundance ratios (Fe/Si and Mg/Si) is reflected in the bulk composition of the majority of super-Earths.  By using this assumption, we can calculate possible interior end-members that account for the anticipated variability of super-Earth structures and compositions.  Furthermore, thermal states of super-Earths are expected to be highly variable, since observed planets have different ages. However, the thermal states of exoplanets are extremely difficult to constrain by observations. We thus use theoretical considerations to account for reasonable ranges of thermal parameters.
On this basis, we investigate and compare how volcanic activity and outgassing is affected by the variability in structural, compositional, and thermal parameters. 

Outgassing is dependent on the convection regime of a planet. It is a matter of debate, what the most likely convection regime is of super-Earths. Here, we focus on the stagnant-lid convection regime in order to fully investigate all relevant parameters, and also briefly discuss other regimes. Furthermore, we restrict the volcanic outgassing to pure CO$_2$, \rev{since it is one of the major outgassed volatiles} \citep{gaillard2014theoretical}. We focus on the accumulated amount of outgassed CO$_2$ over the  lifetime of 4.5 Gyr in order to compare with Solar System planets.  In addition, we discuss the time dependence of outgassing for a range of planet masses and \rev{for ages up} to 10 Gyrs (see Section \ref{Atimedep}).

The paper is structured as follows.
We first provide an introduction on convection regimes and previous studies, we then describe our methodology and results based on the large number of planet simulations. We provide scaling laws for parameters of first and second order influence and end with discussion and conclusions.

\subsection{{Convection regimes}}

For a rocky planet there are commonly three possible convection regimes considered in geodynamics: mobile lid (potentially resembling plate tectonics \citep{mallard16}), stagnant-lid \citep{solomatov95} and episodic regime \citep{moresi98,stein04}. Yet, new convection regimes based on thermo-compositional convection are being reported \citep{moore13,gerya14,sizova15,fischer2016early,lourenco16, rozel2017continental}. 

The stagnant-lid case is characterized by a very rigid lithosphere,  which naturally occurs \rev{when} lithospheric deformation is incapable of triggering mechanisms for localizing shear and weakening the high viscosity lithosphere. In this case, the resistance of rocks to deform in the presence of stress, i.e. the viscosity, is high.   If no other rheological mechanism is included, the lithosphere is so viscous that in cannot be recycled in the deep mantle \citep{solomatov95} and deformation only occurs in the sublithospheric mantle. In this case, outgassing is possible by eruption of melt.
Mercury, Mars and the Earth's moon are examples of stagnant-lid regimes, for which their very intense craterisation gives evidence that the lithosphere has not been entrained in the deep mantle since billions of years. 

The lithosphere of the Earth does extensively deform due to several complex mechanisms \citep{kohlstedt1995strength}: brittle failure \citep{byerlee1978friction}, evolution of microstructures at plate boundaries \citep{burovplate}, rock hydration-weakening and associated phase transitions \citep{mackwell1998high,schwartz2001numerical}, magmatism \citep{marsh2007magmatism}, etc. Due to a combination of all these processes, deformations of the lithosphere can result in a mobile-lid regime (i.e., plate tectonics) in which the lithosphere is constantly recycled in the mantle. This allows greenhouse gases (e.g., CO$_2$, H$_2$O) to cycle between mantle and atmosphere reservoirs by volcanism and  {subduction of carbonate sediments which result from weathering and erosion of surface rocks.}

 \rev{If mantle driving forces do not exceed lithospheric yield strength}, the lithosphere slowly thickens and stresses grow until lithosphere deformation suddenly occurs through a catastrophic event during which the entire lithosphere sinks in the mantle \citep{fowler85,Reese98} and outgassing is efficient \citep{gillmann2014atmosphere}. In this so-called episodic regime,  lithospheric growth and catastrophic resurfacing events happen episodically \citep{moresi98,stein04}.  Venus might experience similar dynamics \citep{strom94}.

\paragraph{{Likelihood of convection regimes}}
Determining the likelihood of convection regimes for super-Earths is still a very challenging problem in geodynamics.
Many interdependent physical parameters are suspected to have a major effect on the dynamics of the lithosphere, which controls the global behaviour of planetary mantles. The strikingly different regime behaviours between Earth and Venus indicates that other parameters besides planet mass and size are determining factors. The difference in solar incident fluxes is often used to explain their respective convection regimes, however, potential key parameters include rock hydration, thermal state, viscosity, melt fraction, compositional heterogeneities and grain size distributions. Heavy numerical implementations and computational resources are required to test these parameters in order to obtain robust scaling laws for the likelihood of different convection regimes. 

After the discovery of the first exoplanets, different studies estimate the likelihood of plate tectonics with increasing planet mass and conclude increasing \citep{valencia2007inevitability,papuc2008internal,valencia2009convection} and decreasing trends \citep{kite}. Furthermore, effects of rock hydration \citep{korenaga2010likelihood} and thermal states (internal heating versus basal heating and initial temperatures) \citep{van2011plate, noack2014plate}, as well as complex rheologies and the pressure-dependence of many physical quantities \citep{tackley2013mantle} can have first-order influences. 
Overall, the likelihood of different convection regimes for super-Earths is ongoing research. Here, we focus on the stagnant-lid regime only.

\subsection{{Previous studies}}

In the following, we highlight few principle studies that investigated outgassing on stagnant-lid planets.
\citet{kite} predict that stagnant-lid exoplanets have high melting rates even for massive super-Earth planets, but they did not consider that melt may be denser than surrounding solid mantle material at specific depths, leading to gravitationally stable melt, thus hindering surface volcanism and outgassing. Also, they considered a purely temperature-dependent viscosity, which is expected to overestimate the mantle convective velocities, and therefore leads to increased melting rates.

\citet{vilella2017fully} derived improved scaling laws for planets for variable convection strength and predict the thermal evolution and melt occurrence on Earth-like exoplanets. They propose that the occurrence of melting decreases with age and planetary radius. Large planets would only show melting early on in their evolution. This study also does consider gravitationally stable melt. 

\citet{noack2014can} investigated the outgassing efficiency for planets of variable core sizes and fixed Earth-like composition and size. Outgassing is strongly reduced for large core radius fractions ($>$0.7 \RE) due 
to the larger pressure gradient in the lithosphere. However, how likely such large core radius fractions are among super-Earths requires further research. While varying Earth-like planets to masses of up to 10 \ME assuming magnesium-silicate mantles and different core-mass fractions, \citet{noack17melting} find that outgassing is limited to planets below 4-7 \ME (depending on other parameter assumptions).

 {Our study differs in several respects compared to the previous study of \citet{noack17melting}:
\begin{itemize}

    \item We test an extensive range of parameters for their influence on mantle outgassing, including planet mass, radiogenic heating, initial mantle temperature, initial lithosphere thickness, mantle composition in terms of Mg/Si and Fe/Si, viscosity, density-cross-over pressure, and effects of hydration.
    \item The range of tested parameters reflect our anticipated variability of the majority of exoplanet interiors.
    \item Our planet interior model allows for general mantle compositions in the FeO-SiO-MgO system.
    \item We quantify the influence of individual parameters on outgassing by providing a scaling law.
\end{itemize}
}

\section{Methodology}
\label{Methodology}

\subsection{Calculation of interior end-members}\label{sec:model_CD}

 The first part of this study concerns the calculation of interiors that cover the anticipated variability of super-Earths. We calculate those interiors given commonly observed ranges of astrophysical data and theoretical prior considerations. The astrophysical data include planetary mass and radius, stellar bulk abundances, and associated uncertainties (listed below). Chosen data uncertainties compare to high data quality. For a specific super-Earth case, we use the probabilistic method of \citet{dorn2015can} to calculate the possible range of interiors. From this range, we identify those interiors of minimum and maximum core size that fit data within 1-$\sigma$ uncertainty. These represent the extracted end-members, which are input to the convection model. The extracted models provide profiles for temperature, density, thermal expansion coefficient, thermal heat capacity, thermal conductivity, gravity, and pressure.
 We provide more details on data and interior model in the following and refer to \citet{dorn2015can} for more details on the probabilistic method.
 
 \paragraph{Data}
 The considered astrophysical data comprise the following, which are listed in Table \ref{tab:data1} and illustrated in Figure \ref{fig:MR} and \ref{fig:Ratios}:
 \begin{itemize}
     \item planetary mass $M_{\rm p}$ (Table \ref{tab:data1}, uncertainty is fixed to 10 \%),
     \item planetary radius $R_{\rm p}$ (Table \ref{tab:data1}, uncertainty is fixed to 5 \%),
     \item bulk abundance $\fesi$ (Fe/Si$_{\rm bulk}$ = \{0.5, 1., 1.5\} $\times$ Fe/Si$_{\rm Sun}$, see Figure \ref{fig:Ratios}, uncertainty is fixed to 20 \%),
     \item bulk abundance $\mgsi$ (Mg/Si$_{\rm bulk}$ = \{0.5, 1., 1.5\} $\times$ Mg/Si$_{\rm Sun}$, see Figure \ref{fig:Ratios}, uncertainty is fixed to 20 \%),
     \item surface temperature is set to 280 K for all cases. 
 \end{itemize}
 {Masses and radii are chosen such that they follow the mass-radius relationship of Earth-like interiors} \citep{valencia2007detailed}.

 \begin{table}[ht]
\caption{Summary of planetary mass and radius data. Uncertainties on mass and radius are 10\% and 5\% , respectively. \label{tab:data1}}
\begin{center}
\begin{tabular}{cc}
\hline\noalign{\smallskip}
{$M_{\rm p}$/\ME}& {$R_{\rm p}$/\RE}   \\
\noalign{\smallskip}
\hline\noalign{\smallskip}
1.& 1.\\
1.5& 1.1\\
2.& 1.2\\
2.5& 1.28\\
3.& 1.33\\
4.& 1.44\\
5.5& 1.52\\
6.6& 1.6\\
7.7& 1.69\\
8.8& 1.74
\end{tabular} 
\end{center}
\end{table}
 
\paragraph{Interior model}

Our planet interior model consists of a layered sphere with an iron core surrounded by a silicate mantle. We allow for variable mantle composition and thicknesses of core and mantle. For the mantle composition, we use the FMS model chemical system that comprises the oxides FeO--MgO--SiO$_2$. Thus the interior parameters comprise:
\begin{itemize}
    \item core size $r_{\rm core}$,
    \item size of core and mantle $r_{\rm core+mantle}$,
    \item $\fesima$,
    \item $\mgsima$.
\end{itemize}

The prior distributions for the model parameters are stated in Table \ref{tableprior} and are similar to those in \citet{dorn2015can,dorn2017generalized}.

 \begin{table}[h]
\caption{Prior ranges.  \label{tableprior}}
\begin{center}
\begin{tabular}{p{.085\textwidth}p{3.13cm}p{3.3cm}}
\hline\noalign{\smallskip}
parameter & prior range & distribution  \\
\noalign{\smallskip}
\hline\noalign{\smallskip}
$r_{\rm core}$ & (0.01  -- 1) $r_{\rm core+mantle}$ &uniform in $r_{\rm core}^3$\\
$r_{\rm core+mantle}$   & (0.01 -- 1) $R_{\rm p}$& uniform in $r_{\rm core+mantle}^3$\\
$\fesima$           & 0 -- $\fesi$&uniform\\
$\mgsima$         & $\mgsi$ &Gaussian
\end{tabular} 
\end{center}
\end{table}
 
We calculate the interiors using self-consistent thermodynamics for core and mantle. For the core  we use the equation of state (EoS) fit of iron in the hcp (hexagonal close-packed) structure provided by \citet{bouchet2013ab} on {\it ab initio} molecular dynamics simulations.
For the silicate mantle, we compute equilibrium mineralogy and density as a function of pressure, temperature, and bulk composition by minimizing Gibbs free energy \citep{connolly2009geodynamic}. We assume an adiabatic temperature profile for core and mantle. 

The interior model is used to calculate interior end-members for super-Earths, that are subsequently used as input to the convection model in order to study melting and outgassing.

\subsection{Convection and melting model}\label{convection}

 The employed convection and melting model is described in detail by \citet{noack17melting}, but briefly outlined in the following.
 
We model convection in a compressible mantle in the 2-D spherical annulus geometry \citep{hernlund2008modeling}. In order to describe compressible flow, we  use the truncated anelastic liquid approximation (TALA). In this approximation, radial reference profiles are used together with calculated lateral variation fields for temperature, density, and pressure \citep[e.g.][]{schubert2001mantle,king2010community, noack17melting}. The reference profiles are those of temperature, density, gravity, and pressure, as well as material properties of thermal expansion coefficient, thermal heat capacity, and thermal conductivity. These profiles are provided by the extracted end-member interiors (see Section \ref{sec:model_CD}). Given the TALA formulation,
the convection code solves the conservation equations for mass, momentum and energy \citep{king2010community,noack17melting}.

The convection behaviour of the mantle depends on the rheological properties of the generally polycrystalline rocks. Here, we use rheology laws that were developed specifically for Earth's mantle. For pressures in the upper mantle,
we use the  {diffusion} 
law for dry olivine from \citet{karato1993rheology}, 
 {
\begin{equation}
\eta(T,p) =  2.6 \cdot 10^{10} \exp{\left( \frac{3\cdot 10^5+ 6\cdot 10^3 p}{RT} \right)} ,\nonumber
\end{equation}
using the universal gas constant $R$;} for pressures in the lower mantle, we use those of perovskite  {(pv)} and post-perovskite  {(ppv)} as derived by \citet{tackley2013mantle} {,
\begin{equation} 
\begin{array}{ll}
\eta(T,p) =  2.5 \cdot 10^{11} \exp{\left( \frac{3.7\cdot 10^5+ 3.65 \cdot 10^3 \exp(\frac{-p}{200}) p}{RT} \right)} & \mbox{, for pv} \\
\eta(T,p) =  3.6 \cdot 10^{8} \exp{\left( \frac{7.8\cdot 10^5+ 1.7 \cdot 10^3 \exp(\frac{-p}{1100}) p}{RT} \right)} & \mbox{, for ppv}
\end{array}\nonumber
\end{equation} 
for pressure $p$ given in GPa and temperature $T$ in K.}
Thereby, we neglect compositional effects on rheology.  {However, we do investigate the role of the viscosity on our outgassing results by adding a viscosity prefactor $\Delta_\eta$ which is set to $10$ in case 9 and $1$ in all other cases. For the rheology laws given above, we obtain a reference viscosity of $1.6\cdot 10^{20}$, $3\cdot 10^{23}$ and $1\cdot 10^{34}$ Pa\ s for olivine, perovskite and post-perovskite, respectively, at a reference temperature of $1600$ K and zero GPa.} 

Melting is tracked at every time step in our simulations. Where mantle temperature exceeds the solidus temperature, partial melting occurs. If the melt is gravitationally buoyant, we assume that melt should rise immediately to the surface and outgas. Instead of transporting the melt to the surface, we calculate the amount of CO$_2$ that should be outgassed. The residue is consequently depleted in volatiles. We use the same parametrization as in \citet{noack17melting} for outgassing processes (Table \ref{tab:refCase}), i.e., if melting occurs at pressures below the so-called density cross-over pressure ($P_{\rm cross-over}$), the melt with initially 1000 ppm of CO$_2$ rises to the surface and depletes by 10\% in volatiles. Depending on mantle mixing and the occurrence of partial  melting, this process can happen repeatedly, however, maximum mantle depletion $d_{\rm max}$ is set to 30\% (volumetric fraction). Mantle depletion is thus directly linked to the amount of outgassed volatiles. To trace the volatile depletion in the mantle, we use a particle-in-cell approach.

\begin{figure}
   \centering
   \includegraphics[width=1\hsize]{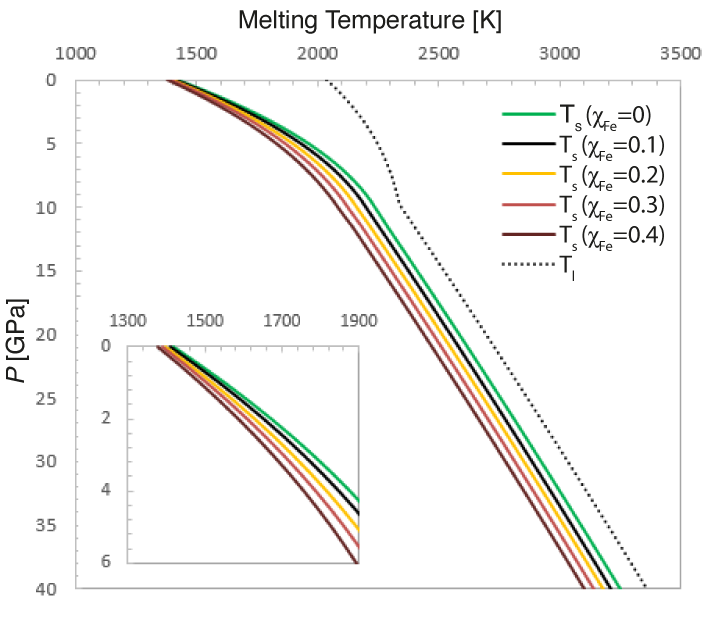}
   \caption{Earth-like solidus ($T_{\rm s}$, black solid line) and liquidus ($T_{\rm l}$, black dotted line) in comparison with melting temperatures as a function of iron weight fraction of the mantle ($\mathcal{X}_{Fe}$).}
              \label{fig:melt_iron}
    \end{figure}

The amount of outgassing can be affected by mantle composition, because melting temperatures depend on rock composition \citep{kiefer2015melt}.
Based on the laboratory studies summarised in \citet{kiefer2015melt} and \citet{hirschmann2000melt}, 
we derive an iron-dependent melting law for low pressures. This is an addition to the usual  solidus temperatures. For pressures above 12~GPa, the iron influence on the melting temperature is assumed to be pressure-independent:
\begin{equation} 
\begin{array}{ll}
\Delta_{T_{s,Fe}} = (102 + 64.1 p - 3.62 p^2)\cdot(0.1-\mathcal{X}_{Fe}) & \mbox{, if $p\le$12} \\
\Delta_{T_{s,Fe}} = 360\cdot (0.1-\mathcal{X}_{Fe}) & \mbox{, else.}
\end{array}\nonumber
\end{equation} 
The iron content $\mathcal{X}_{Fe}$ is given in mass fraction and the pressure $p$ in GPa. The melting temperature for iron contents between 0 and 0.4 is depicted in Fig. \ref{fig:melt_iron}.

For some test cases (10 and 11), we account for hydrated rock and use a wet solidus  {formulation taken from \citet{katz2003new} by assuming an initial amount of 500 wt-ppm water in all mantle rocks. The influence of water on the solidus is
\begin{equation}
\Delta_{T_{\rm s,H2O}} = -43 \mathcal{X}_{\rm H2O}^{0.75},\nonumber
\end{equation}
where water content here is in wt-\%. Due to partial melting, water partitions into the melt for small melting degrees, and the residual material is set as dehydrated for melting depletion above 5 wt-\%.}
The melting solidus and liquidus temperatures  {for Earth-like mantle iron content $T_s$ and $T_l$} are taken from \citet{hirschmann2000melt}. 
 {The effective solidus temperature is then calculated as
\begin{equation}
T_{\rm s,eff} = T_{s} + \Delta_{T_{\rm s,Fe}} + \Delta_{T_{\rm s,H2O}}.\nonumber
\end{equation}
}
Due to lack of experimental data, a more detailed treatment of the influence of composition on melting temperature is not justified for our study.

 {Initial temperatures in the mantle are cut if they lie above the solidus temperature, to avoid initial melting induced purely by the initial setup of the mantle. However, to be able to compare the simulations with a wet and dry solidus, for our wet mantle cases, we cut the initial mantle temperatures only if they lie above the dry solidus temperature.}

 {We model the thermal evolution of all model planets over time. The initial temperature profile is calculated adiabatically starting from an initial upper mantle temperature $T_{\rm init,mantle}$, which is set beneath the lithosphere with an initial thickness of $T_{\rm init,mantle}$. For most cases, we treat the core as isolated from the mantle, which means that the mantle temperatures evolve solely depending on the heat flux through the lithosphere, radioactive heat sources in the mantle, and latent heat consumption by melting. No heat flux from the core into the mantle is considered. In case 7, instead, we assume at the core mantle boundary an initial temperature difference between mantle and core $\Delta T_{\rm cmb}$ scaled with planet mass \citep{stixrude2014melting}. For this test case, the core cools with time and adds as additional heat source for the mantle. Radioactive heat sources are varied between the different cases from 0.5 to 1.5 times Earth-like initial amount of heat sources, and decay over time (see Table \ref{tab:refCase}).  {For} Earth-like initial mass concentration of radiogenic elements  {we assume} 
c$_{\rm U^{235}}$ = 1.2$\times10^{-8}$, c$_{\rm U^{238}}$ =   4.0$\times10^{-8}$, c$_{\rm Th^{232}}$ = 9.9$\times10^{-8}$, c$_{\rm K^{40}}$ =  3.7$\times10^{-7}$. At 4.5 Gyrs, these mass concentrations are c$_{\rm U^{235}}$ = 1.4$\times10^{-10}$, c$_{\rm U^{238}}$ =   2.0$\times10^{-8}$, c$_{\rm Th^{232}}$ = 7.9$\times10^{-8}$, c$_{\rm K^{40}}$ =  3.1$\times10^{-8}$ \citep{mcdonough1995composition} and are summarized as 1 \CE. The total radiogenic heat production rate at 0 Gyrs is 24.2 pW/kg.}

\section{Results}
\label{Results}

We compiled a set of  {2340} super-Earth models, for which we study melting and outgassing. This set  covers our anticipated diversity of super-Earths in terms of structural, compositional, and thermal parameters. 

 {The set comprises super-Earths of 10 different masses and radii (see Table \ref{tab:data1}) and 9 different bulk composition constraints (see Figure \ref{fig:Ratios}), 6 different thermal parameters that stem from different formation conditions, and 7 other parameters relevant for melting and interior dynamics} (see Table \ref{tab:refCase}).  {This yields a total of $10\times9\times13 = 1170$ models}. In addition, for each super-Earth model we consider two interior end-members (i.e., the models with minimum and maximum core size that fit data constraints). Thus, we have a total of  {$1170\times2 = 2340$} super-Earth models, for which we simulate outgassing over a lifetime of 4.5 Gyr  {(see Section \ref{Atimedep} for time-dependence of outgassing up to 10 Gyrs)}. {The reference case comprises $10\times9\times2 = 180$ models as shown in Figure \ref{fig:mass}}.
In the following, we discuss the individual cases.

 \begin{table*}[ht] \setlength{\tabcolsep}{5pt}
\caption{Input parameters of considered test cases, where $Q_{\rm rad}$ are amounts of radioactive heat sources, $T_{\rm init,mantle}$ is the initial upper mantle temperature, $D_{\rm init,lith}$ is the initial lithosphere thickness, $\Delta T_{\rm cmb}$ is the temperature jump at the core-mantle-boundary (CMB),  {and $P_{\rm cross-over}$ is the density-cross-over pressure}. $Q_{\rm rad}$ is in units of \CE that is the Earth-like amounts of radioactive heat sources and represent present-day values \citep{mcdonough1995composition}, from which initial amounts 4.5 billions years ago are calculated. Bold values indicate a variation with respect to the reference case. \label{tab:refCase}}
 \widesplit{%
\begin{tabular}{l|*{13}{l}}
\hline\noalign{\smallskip}
{\bf Parameter} \row{& \bf Reference & \bf Case 2 & \bf Case 3 & \bf Case 4 & \bf Case 5 & \bf Case 6 & \bf Case 7 }{& \bf Case 8 & \bf Case 9 & \bf Case 10 &\bf  Case 11 &\bf  Case 12 &\bf  Case 13} \\
\hline\noalign{\smallskip}
$Q_{\rm rad}$ \row{& 1\CE & {\bf 1.5\CE}& {\bf 0.5\CE}&1\CE &1\CE &1\CE &1\CE }{&1\CE &1\CE &1\CE &1\CE &1\CE &1\CE}\\
$T_{\rm init,mantle}$ \row{& 1800 K & 1800 K& 1800 K& {\bf 1600 K}& {\bf 2000 K} &1800 K & 1800 K}{&1800 K&1800 K&1800 K&\bf 1600 K&1800 K&1800 K}\\
$D_{\rm init,lith}$ \row{& 100 km & 100 km&100 km &100 km &100 km &{\bf 50 km} & 100 km}{&100 km&100 km&100 km&100 km&100 km&100 km}\\
$\Delta T_{\rm cmb}$ at the CMB \row{& 0 K & 0 K& 0 K& 0 K& 0 K& 0 K& {$\bf \Delta \bf T_{\rm cmb}(M_{\rm p}$/\ME)}$^*$}{& 0 K& 0 K& 0 K& 0 K& 0 K& 0 K}\\
Radial grid resolution  \row{& 25 km  & 25 km &25 km &25 km  &25 km  &25 km  &25 km  }{&{\bf 10 km} &25 km&25 km&25 km&25 km&25 km}\\
viscosity prefactor $\Delta_\eta$ \row{&1 &1 &1 &1 &1&1&1}{&1 &\bf 10&1&1&1&1}\\
Wet/dry solidus  \row{& dry& dry &dry &dry  &dry  &dry &dry }{& dry & dry &\bf wet &\bf wet&dry&dry}\\
$P_{\rm cross-over}$ \row{& 12 GPa& 12 GPa&12 GPa &12 GPa  &12 GPa  &12 GPa &12 GPa }{& 12 GPa& 12 GPa &12 GPa &12 GPa&\bf 8 GPa&\bf 16 GPa}\\
\hline\noalign{\smallskip}
Surface temperature  \row{& \multicolumn{7}{c}{280 K}}{& \multicolumn{6}{c}{280 K}}\\
Particles per cell \row{& \multicolumn{7}{c}{10}}{& \multicolumn{6}{c}{10}}\\
Latent heat  \row{& \multicolumn{7}{c}{600 kJ/kg}}{& \multicolumn{6}{c}{600 kJ/kg}}\\
Max. mantle depletion $d_{\rm max}$ \row{& \multicolumn{7}{c}{30 \%}}{& \multicolumn{6}{c}{30 \%}}\\
Amount of CO$_2$ in melt $f_{\rm CO_2}$ \row{& \multicolumn{7}{c}{1000 ppm}}{& \multicolumn{6}{c}{1000 ppm}}\\
Extrusive volcanism $f_{\rm ex}$ \row{& \multicolumn{7}{c}{10 \%}}{& \multicolumn{6}{c}{10 \%}}\\
time of evolution \row{& \multicolumn{7}{c}{4.5 Gyr}}{& \multicolumn{6}{c}{4.5 Gyr}}\\
\hline
 \multicolumn{5}{l}{\footnotesize{$^*$ The function $\Delta T_{\rm cmb}$($M_{\rm p}$ /\ME) =  1400 K ($M/$\ME)$^{3/4}$ is taken from \citet{stixrude2014melting}.}}
\end{tabular}%
}
\end{table*}

\subsection{Outgassing versus planet mass}
\label{versusmass}

   \begin{figure*}
   \centering
   \includegraphics[width=1\hsize]{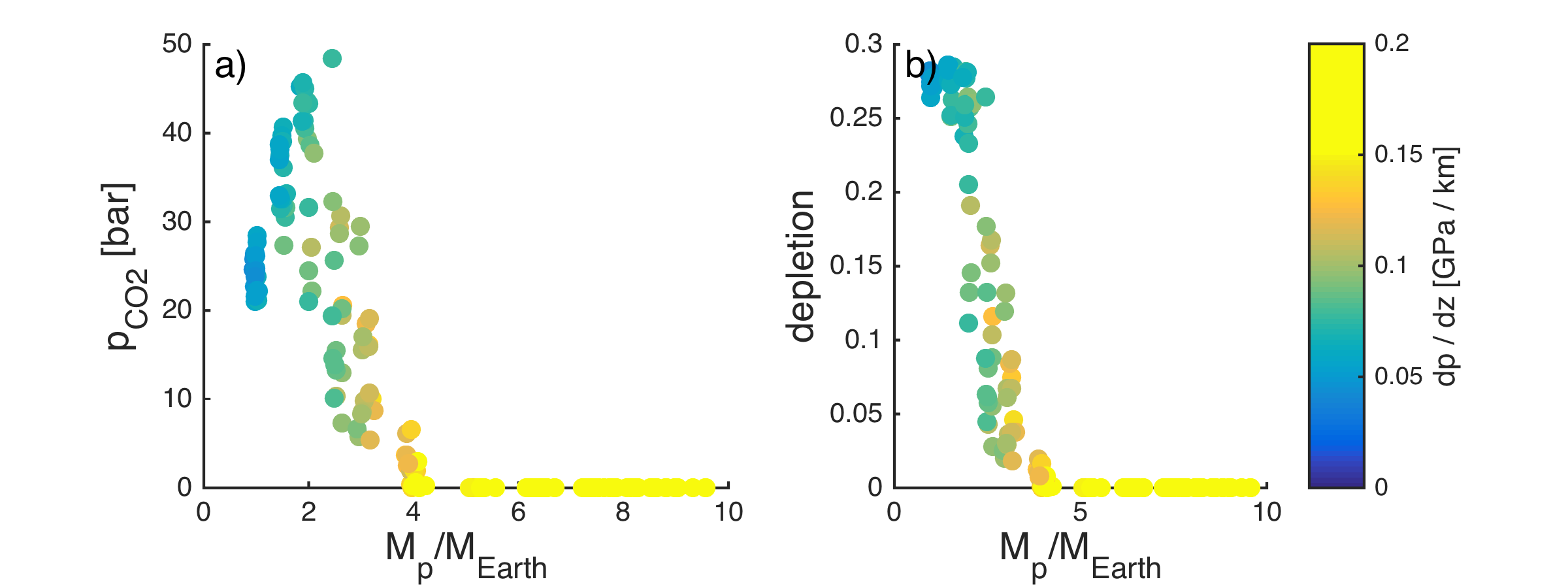}
   \caption{Influence of planet mass $M_{\rm p}$ on (a) outgassing and (b) mantle depletion for the reference case \rev{(180 Super-Earths shown)} after 4.5 Gyr (see Table \ref{tab:refCase}). The amount of outgassing of $\rm CO_2$ is denoted in terms of partial pressure $p_{\rm CO_2}$. Pressure gradient \pg  is shown in color.  {For plotting purposes, we saturated the colorscale at high values of \pg.}}
              \label{fig:mass}
    \end{figure*}


   \begin{figure}
   \centering
   \includegraphics[width = .5\textwidth, trim = 0cm 0cm 0cm 0cm, clip]{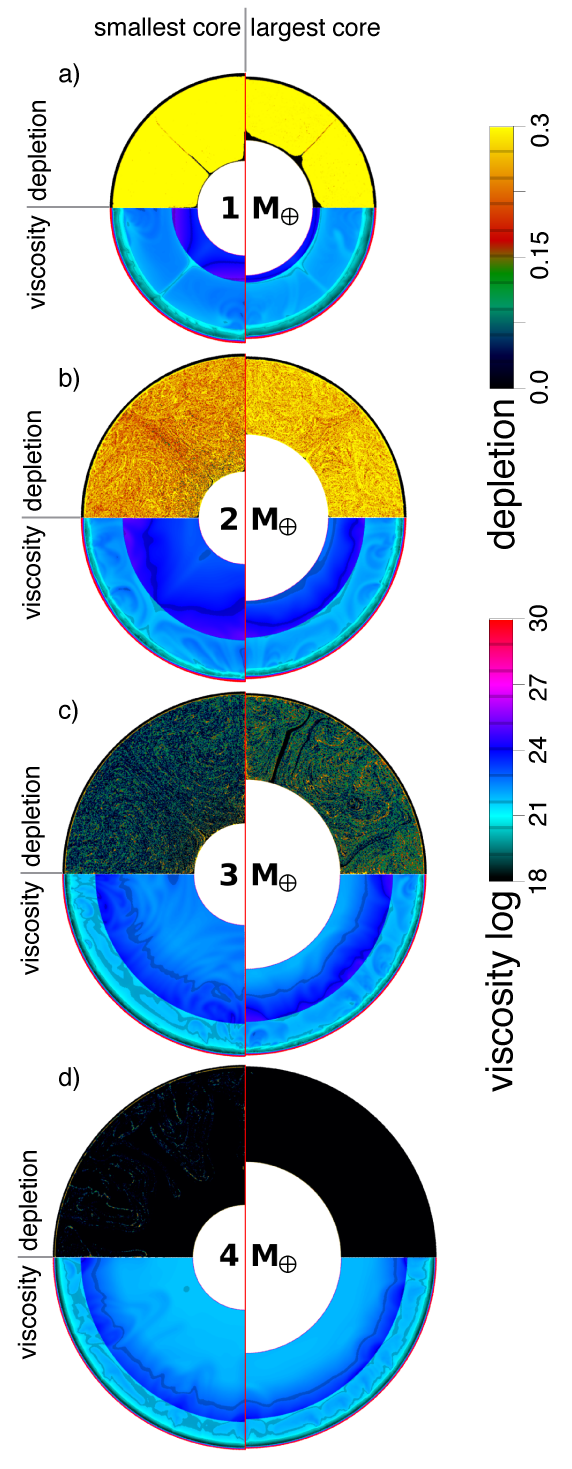}
   \caption{Mantle depletion and viscosity for the reference case and for planets of (a) 1~\ME, (b) 2~\ME, (c) 3~\ME, (d) 4~\ME after 4.5 Gyr. The left and right panels represent smallest and largest core sizes, respectively, that are in agreement with planet bulk abundances ($\fesi$ =$1\times \fesisun$ and $\mgsi$ =$1\times \mgsisun$).}
              \label{fig1}
    \end{figure}

Mantle depletion decreases with larger planet mass $M_{\rm p}$ (Fig. \ref{fig:mass}). For the 1~\ME planet, the mantle is almost completely depleted after 4.5 Gyrs, whereas for planets of 2-4~\ME depletion is significantly reduced (see also Fig. \ref{fig1}). 
This is because at higher masses, the pressure gradient in the lithosphere increases and thereby reduces the depth (or pressure) range, where melting can occur and melt is buoyant. Also, an increasing pressure at the bottom of the lithosphere results in higher melting temperature.  {We note that the pressure gradient \pg that is plotted in Figures \ref{fig:mass} and later} is defined as bulk density ${\rho_{\rm bulk}}$ times gravity (\pg = $g\cdot\rho_{\rm bulk}$) and thus \pg$\sim M_{\rm p}^2/R_{\rm p}^5$. Here, the considered super-Earths roughly follow $R_{\rm p} = M_{\rm p}^{0.26}$ \citep{valencia2007detailed}, thus \pg changes nearly linearly with planet mass.

 {The amount of outgassed volatiles is denoted in partial pressure $p_{\rm  CO_2}$ in bar, which is the mass of outgassed CO$_2$ ($m_{\rm  CO_2}$) times gravity divided by surface area:
\begin{equation}
    p_{\rm  CO_2}  = m_{\rm  CO_2} g / 4 \pi R_p^2 \,.
\end{equation}
The influence of planet mass on $p_{\rm  CO_2}$ is shown in Fig. \ref{fig:mass}.} In this case, the absolute amount of outgassed $\rm CO_2$ increases with planet mass, because the absolute volume  of mantle material and thus the volume of melt is larger. This trend dominates outgassing at small masses (1-2~\ME).

In Figure \ref{fig1} (lower half of each subplot), we show viscosity fields for the reference case at four different masses. Phase transitions between perovskite (pv) and post-perovskite (ppv) in the mantle are visible where viscosity increases by $\sim$ 1-2 orders of magnitude. At large masses, mantle viscosities become relatively uniform which is due to a self-regulatory process \citep{tackley2013mantle}. This process can be understood as follows. Viscosity increases with pressure which tends to decrease the convective vigor. However, this leads to a higher internal temperature of the mantle. Since viscosity is temperature-dependent, viscosities are consequently lowered back to a level where global scale convection occurs.

\subsection{Outgassing versus thermal state}



   \begin{figure*}
   \centering
   \includegraphics[width=1\hsize]{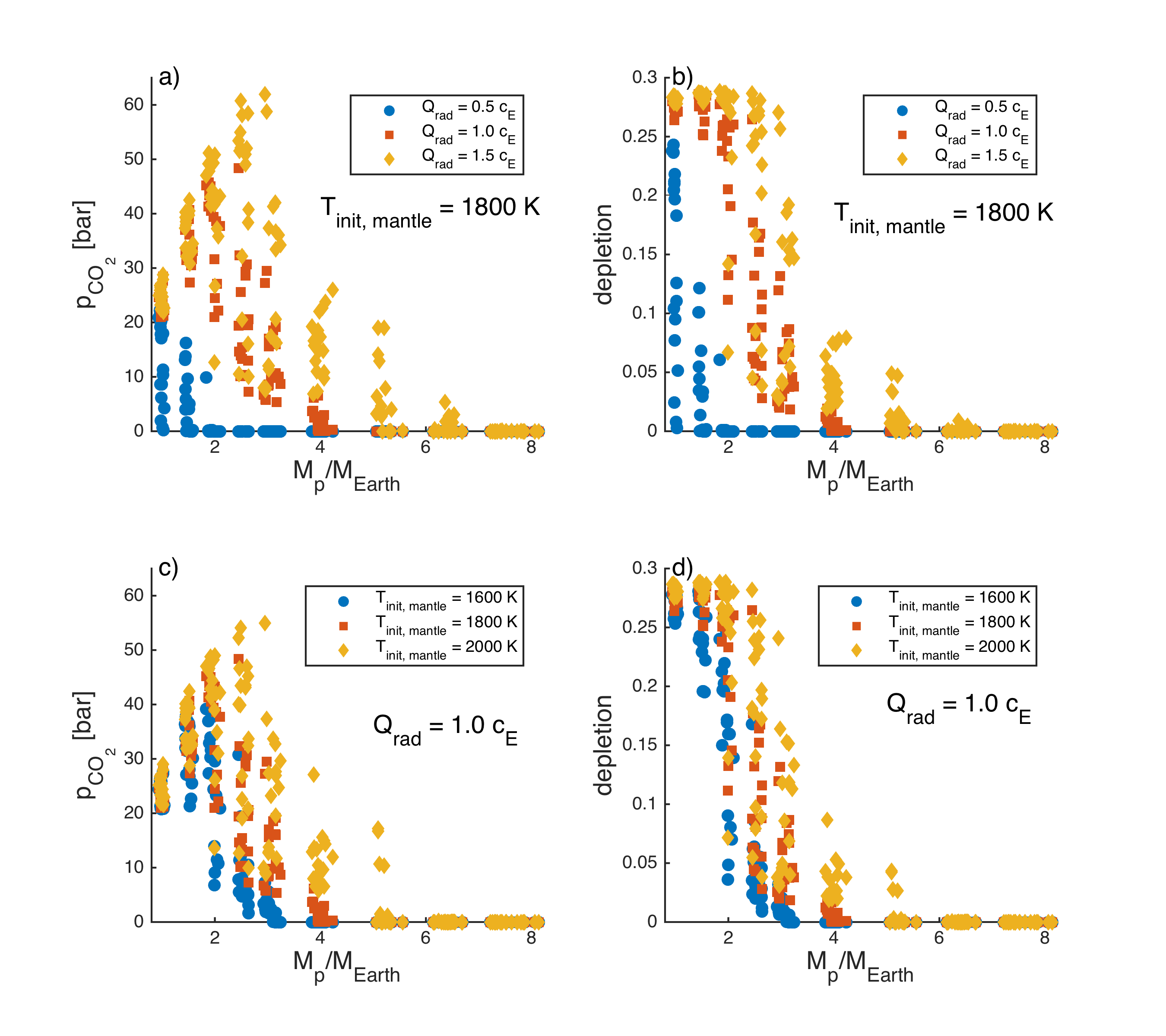}
   \caption{Influence of (a-b) radiogenic heating $Q_{\rm rad}$ and (c-d) initial mantle temperature $T_{\rm init,mantle}$ on (a,c) outgassing and (b,d) mantle depletion. The amount of outgassing of $\rm CO_2$ is denoted in terms of partial pressure $p_{\rm CO_2}$. The cases 2-5 and the reference case are shown (Table \ref{tab:refCase}). }
              \label{fig:T}
    \end{figure*}
Besides planet mass, thermal parameters have first-order effects on depletion and outgassing. Figure \ref{fig:T} shows how much an increase in radioactive heat sources $Q_{\rm rad}$ and initial upper mantle temperatures $T_{\rm init,mantle}$ leads to enhanced depletion and outgassing. Note that $T_{\rm init,mantle}$ is the initial temperature at the boundary between lithosphere and upper mantle.

We vary the amount of radioactive heat sources from 0.5 to 1.5 times the Earth-like values (cases 2 and 3 in Table \ref{tab:refCase}) to cover largely the expected variability range based on galactic evolution models \citep{frank2014radiogenic}  {with regard to 
stellar ages \citep{silva2015ages}}.
An increase in the amount of radioactive heat sources can significantly enlarge the depth range where melting occurs. Thereby it enlarges the mass range of super-Earths, where depletion and outgassing are efficient. For example, maximal depletion is observed up to 1~\ME for $Q_{\rm rad}$ = 0.5~\CE and 3~\ME for $Q_{\rm rad}$ = 1.5~\CE. 
This suggests that planets that formed early in the galactic history tend to be more depleted, since radiogenic heat sources were more abundant \citep{frank2014radiogenic}.

Similarly, an increase from 1600 to 2000 K for the $T_{\rm init,mantle}$ (cases 4 and 5 in Table \ref{tab:refCase}) extends the mass range of maximum mantle depletion from 1.5 to 2.5~\ME. The chosen range of variability in $T_{\rm init,mantle}$ is based on the expected variation of upper mantle temperatures after the magma ocean state of a rocky planet, which is subject to the mantle composition and  estimates for Earth-like compositions broadly covers 1600 - 2000 K \citep{herzberg2010thermal,jaupart20077}. 

In general, an increase in  thermal parameters (i.e., $T_{\rm init,mantle}$, $Q_{\rm rad}$) enables melting at shallower depths which partly outweighs for the pressure-limited melting depths at higher mass planets.  Thereby,  the mass range where depletion is most efficient can be extended up to 3~\ME. 
However, even on the initially hottest super-Earths (case 2 with $Q_{\rm rad}$ = 1.5~\CE and case 5 with $T_{\rm init,mantle}$ = 2000 K) depletion and outgassing only occurs up to 7~\ME.

   \begin{figure}
   \centering
   \includegraphics[width=1\hsize]{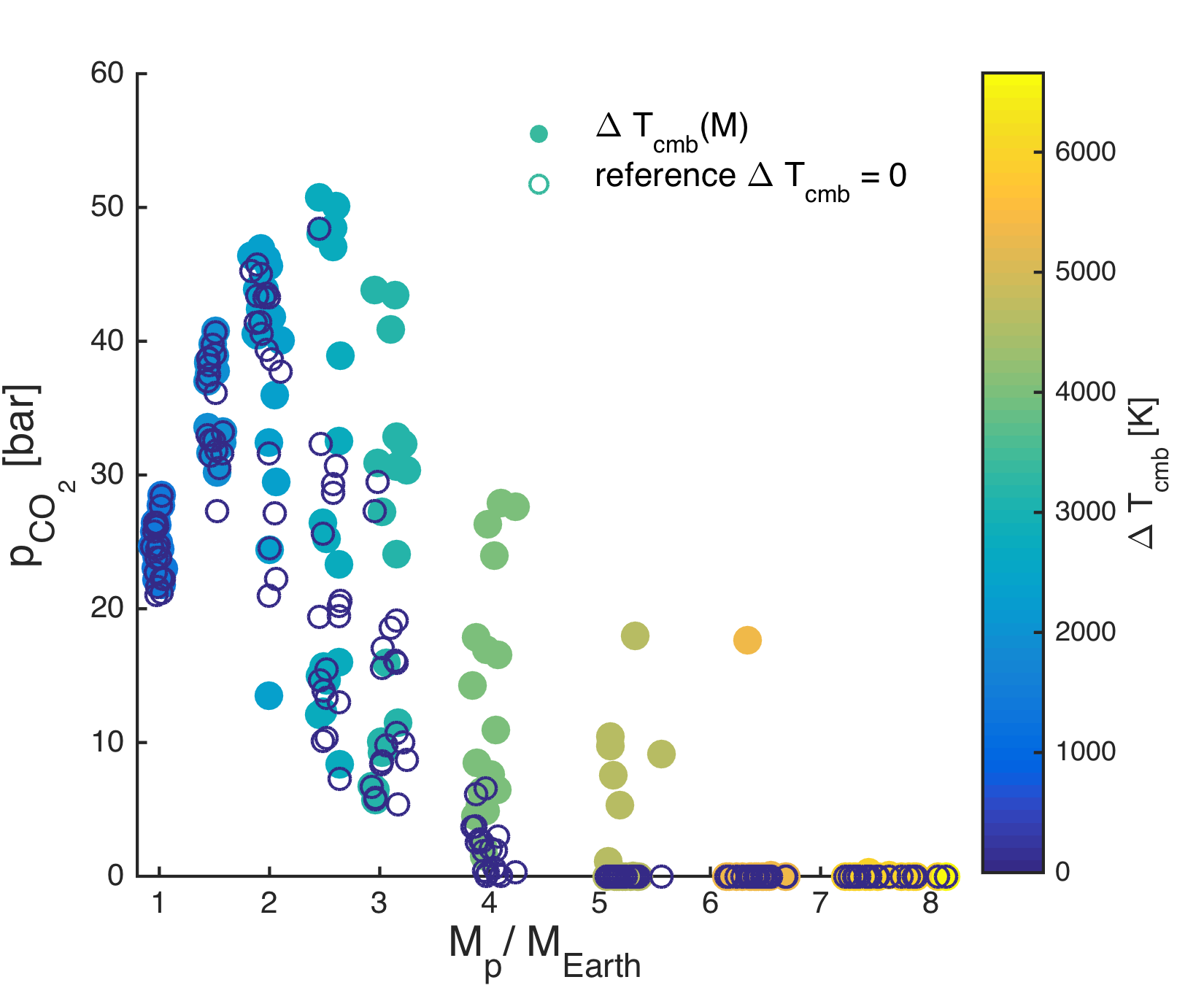}
   \caption{Influence of a temperature increase at the core-mantle boundary $\Delta T_{\rm cmb}$ on outgassing. The amount of outgassing of $\rm CO_2$ is denoted in terms of partial pressure $p_{\rm CO_2}$. The case 7 and the reference case are shown (Table \ref{tab:refCase}).}
              \label{fig:Tcmb}
    \end{figure}

Similar effects are seen when considering super-heated cores that lead to a basally heated mantle. For Earth it is still a matter of debate how much heat flux there is at the CMB and estimates suggest 20\% of the total internal heating \citep{schubert2001mantle}. For the test case 7, we use the mass-dependent power-law
$\Delta T_{\rm cmb}$($M_{\rm p}$ /\ME) =  1400 K ($M/$\ME)$^{3/4}$
by \citet{stixrude2014melting} based on scaled thermal models. 
In general, the overall trend of outgassing on super-Earths is only weakly effected, however, the absolute amounts of outgassing can be significantly higher, especially for high mass planets (3--6~\ME) as shown in Figure \ref{fig:Tcmb}.

\subsection{Outgassing versus interior structure}
\label{interiori}

   \begin{figure*}
   \centering
   \includegraphics[width = .9\textwidth, trim = 0cm 0cm 0cm 0cm, clip]{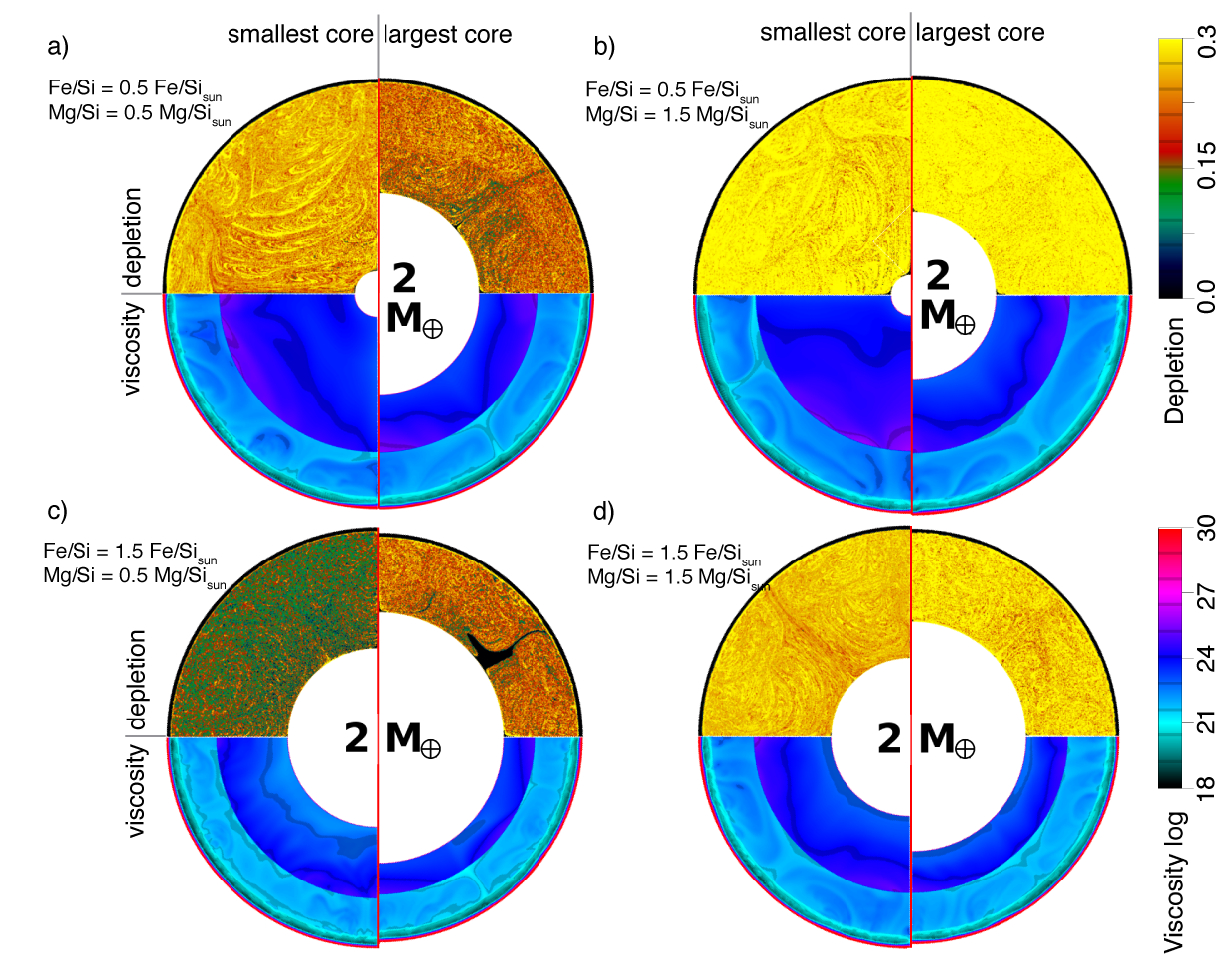}
   \caption{Mantle depletion and viscosity for different planet bulk compositions  after 4.5 Gyr. All planets have a masses of 2~\ME. The left and right panels represent smallest and largest core sizes, respectively, that are in agreement with planet bulk abundances ($\fesi$ and $\mgsi$). The reference case is shown (Table \ref{tab:refCase}).}
              \label{fig2}
    \end{figure*}

The variation of core size and mantle composition seems to have a secondary influence on depletion and outgassing. In Figure \ref{fig1}, we show  planets of similar (solar-like) bulk composition, but with a different distribution of the bulk iron between core and mantle. The planets with large cores (right panels in Fig. \ref{fig1}) have little iron in the mantle, whereas small cores imply a higher iron mantle content. For the planets of 2 and 3 \ME, the interiors with higher iron mantle content and small cores seem less depleted. In this case, the mantle density is higher and leads to a higher pressure at the bottom of the lithosphere which reduces the depth range of buoyant melt production.

In Figure \ref{fig2} for a 2\ME planet, we show the effect on depletion and viscosity due to the variation of bulk composition in terms of $\fesi$ and $\mgsi$. Generally, the planets of high $\mgsi$ and low $\fesi$ tend to be more depleted.
Also, for a given bulk composition, the influence of the core size can result in larger or smaller mantle depletion. In Figure \ref{fig2}c ($\fesi$= 1.5~$\fesisun$ and $\mgsi$= 0.5~$\mgsisun$), a larger core results in higher depletion (similar to Figure \ref{fig1}), whereas for Figure \ref{fig2}a ($\fesi$= 0.5~$\fesisun$ and $\mgsi$= 0.5~$\mgsisun$), we see the opposite. In this case, the reduced mantle depletion can be explained by reduced melting due to an increase in melting temperature with less iron content.

\begin{figure*}
   \centering
   \includegraphics[width=.9\hsize]{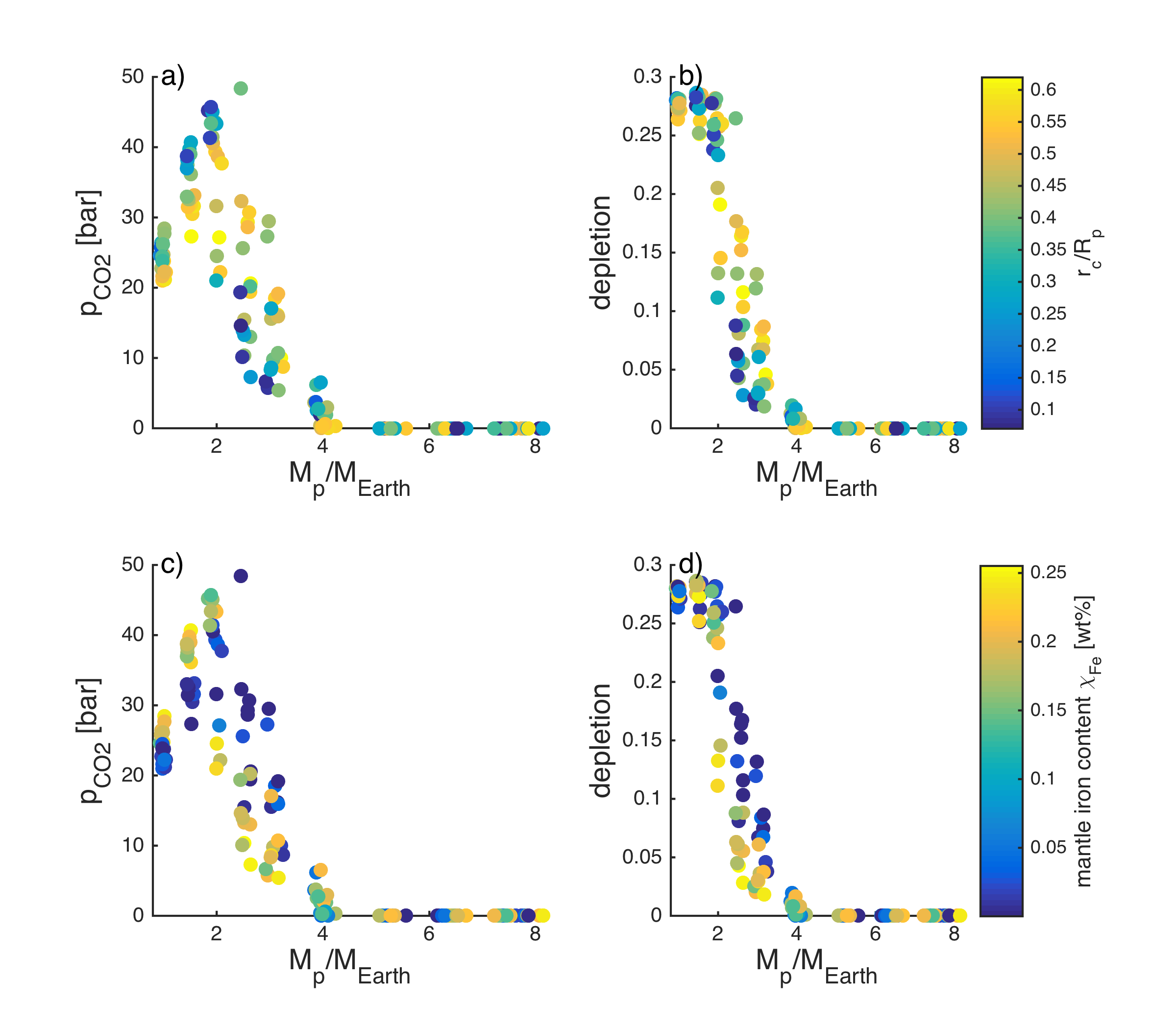}
   \caption{Influence of (a-b) core radius fraction \rc/$R_{\rm p}$  and (c-d) mantle iron content $\mathcal{X}_{Fe}$ on (a,c) outgassing and (b,d) mantle depletion. The amount of outgassing of $\rm CO_2$ is denoted in terms of partial pressure $p_{\rm CO_2}$. The reference case is shown (Table \ref{tab:refCase}).}
              \label{fig:interior}
    \end{figure*}

The dependence of depletion and outgassing on core size and mantle composition is summarized in Figure \ref{fig:interior} for the reference case. For low-mass planets (<~2\ME), we see that increasing core size and decreasing mantle iron content leads to a decreased amount of outgassing. The amount of outgassed volatiles is limited by the absolute mantle volume, i.e., there is less outgassing for large cores. Mantle depletion in these cases is very efficient and weakly dependent on core size and mantle composition. Furthermore, influences of core size and mantle composition seem to become insignificant in the case of low radiogenic heating (case 3 in Table \ref{tab:refCase}) (not shown).

For large-mass planets (>~2\ME), we see the opposite, in that larger core sizes and lower mantle iron contents result in  higher amount of outgassing and higher mantle depletion. 
In these cases, the melting region is relatively shallow and mostly within the lithosphere. High mantle iron contents imply a higher mantle density, which reduces melting by increasing the pressure at the bottom of the lithosphere. Thus melting is reduced to a shallower region. Even though a higher iron content lowers the melting temperature which would imply enhanced depletion, the effect on mantle density is stronger.

\rev{We note that differences in mantle composition affect solidus temperatures as well as the reference profiles of temperature, density, gravity, and pressure, and also material properties of thermal expansion coefficient, thermal heat capacity, and thermal conductivity. Dependencies between composition and viscosity are not taken into account. Instead, we investigate effects of viscosity variations independent of mantle composition, which might overpredict the variability of depletion (see Section \ref{viscoo}).}

\begin{figure}
   \centering
   \includegraphics[width=1\hsize]{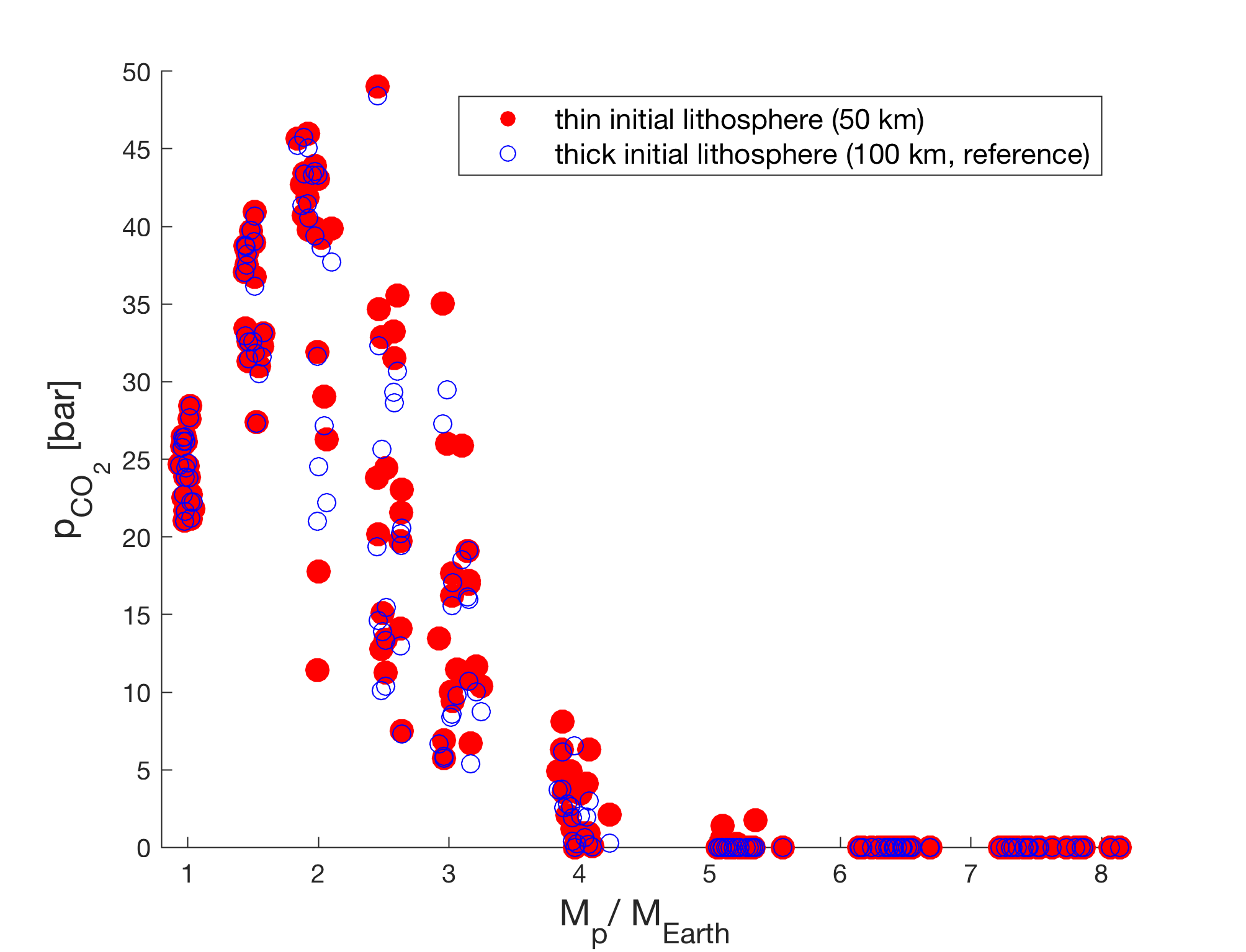}
   \caption{Influence of the initial lithosphere thickness on outgassing. The amount of outgassing of $\rm CO_2$ is denoted in terms of partial pressure $p_{\rm CO_2}$. The case 6 and the reference case are shown (Table \ref{tab:refCase}).}
              \label{fig:Ld}
    \end{figure}

The initial lithosphere thickness (see case 6 in Table \ref{tab:refCase}) only weakly influences volatile outgassing, which is depicted in Figure \ref{fig:Ld} for a thin (50 km) and thick (100 km) initial lithosphere thickness. As expected, a thinner initial lithosphere leads to slightly higher outgassing, since the initial melting depth extends deeper into the mantle.

    \subsection{Outgassing versus viscosity}
\label{viscoo}
   \begin{figure}
   \centering
   \includegraphics[width = .5\textwidth, trim = 0cm 0cm 0cm 0cm, clip]{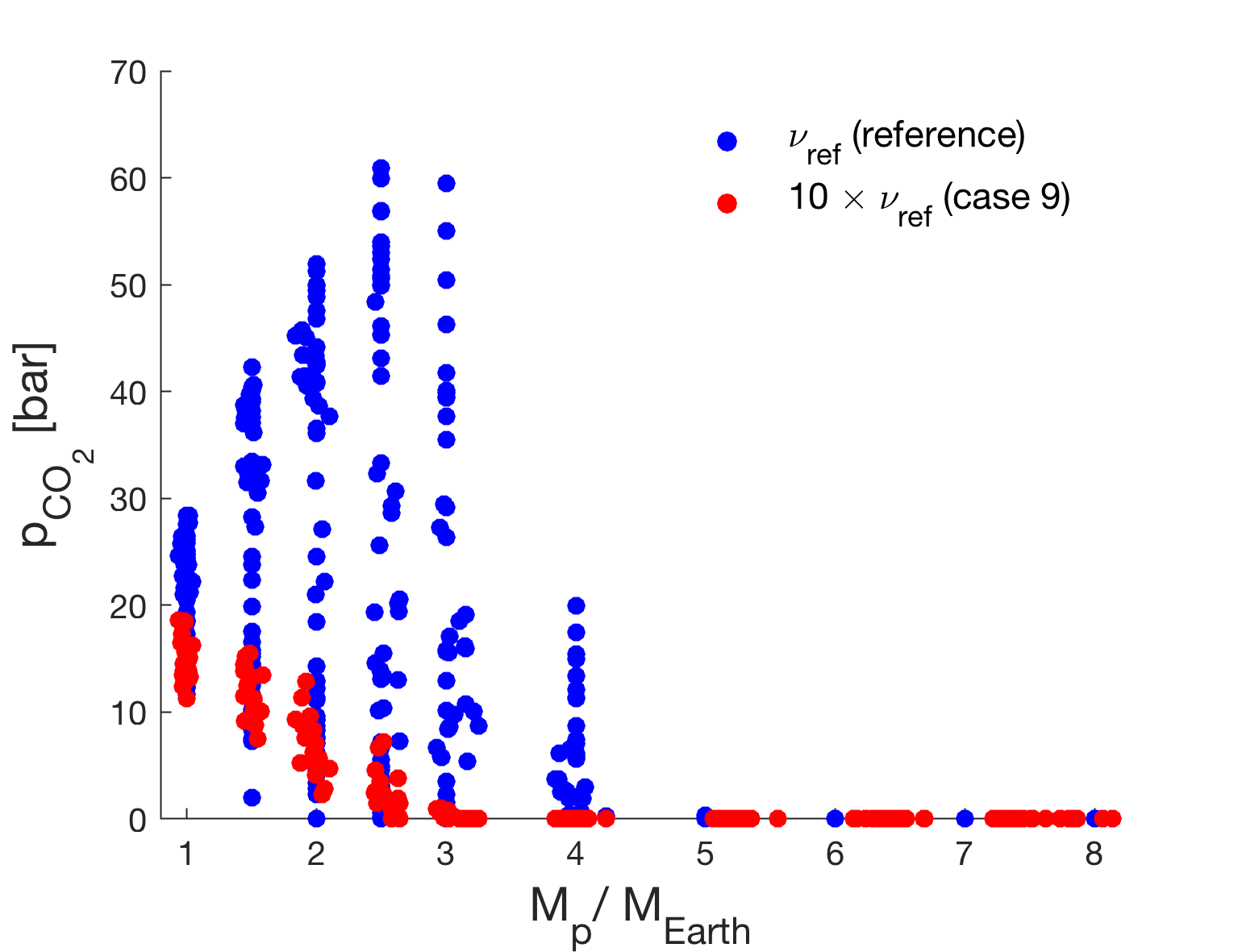}
   \caption{ {Influence of reference viscosity on outgassing as a function of planet mass. Here, we compare the reference case to case 9.}}
              \label{figVIS}
    \end{figure}
     {
    Viscosity can significantly influence
    outgassing. An increased viscosity leads to less vigorous convection, thickens the lid and thereby reduces outgassing. The reduction of outgassing due to an increase in reference viscosity by a factor 10 is shown in Figure \ref{figVIS}. For all planet masses, outgassing is reduced, and the maximal outgassing efficiency is obtained for the smallest investigated mass of 1\ME. A smaller reference viscosity (for example for increased amounts of iron, \citealt{zhao2009effect}, or water, \citealt{HiKo03}, in the mantle) would have the opposite effect.
    
    At low masses ($<$~2\ME), where mantle depletion is most efficient and $p_{\rm CO_2}$ is at its maximum in the reference case, a decrease of viscosity would not further increase $p_{\rm CO_2}$. However, we expect significant influences at intermediate masses (2-4 \ME).
This intermediate mass range is also where the higher viscosity significantly reduces volcanic outgassing.
}
    
        \subsection{Outgassing versus buoyant behaviour of melt}
\label{crossy}
\begin{figure}
   \centering
   \includegraphics[width = .5\textwidth, trim = 0cm 0cm 0cm 0cm, clip]{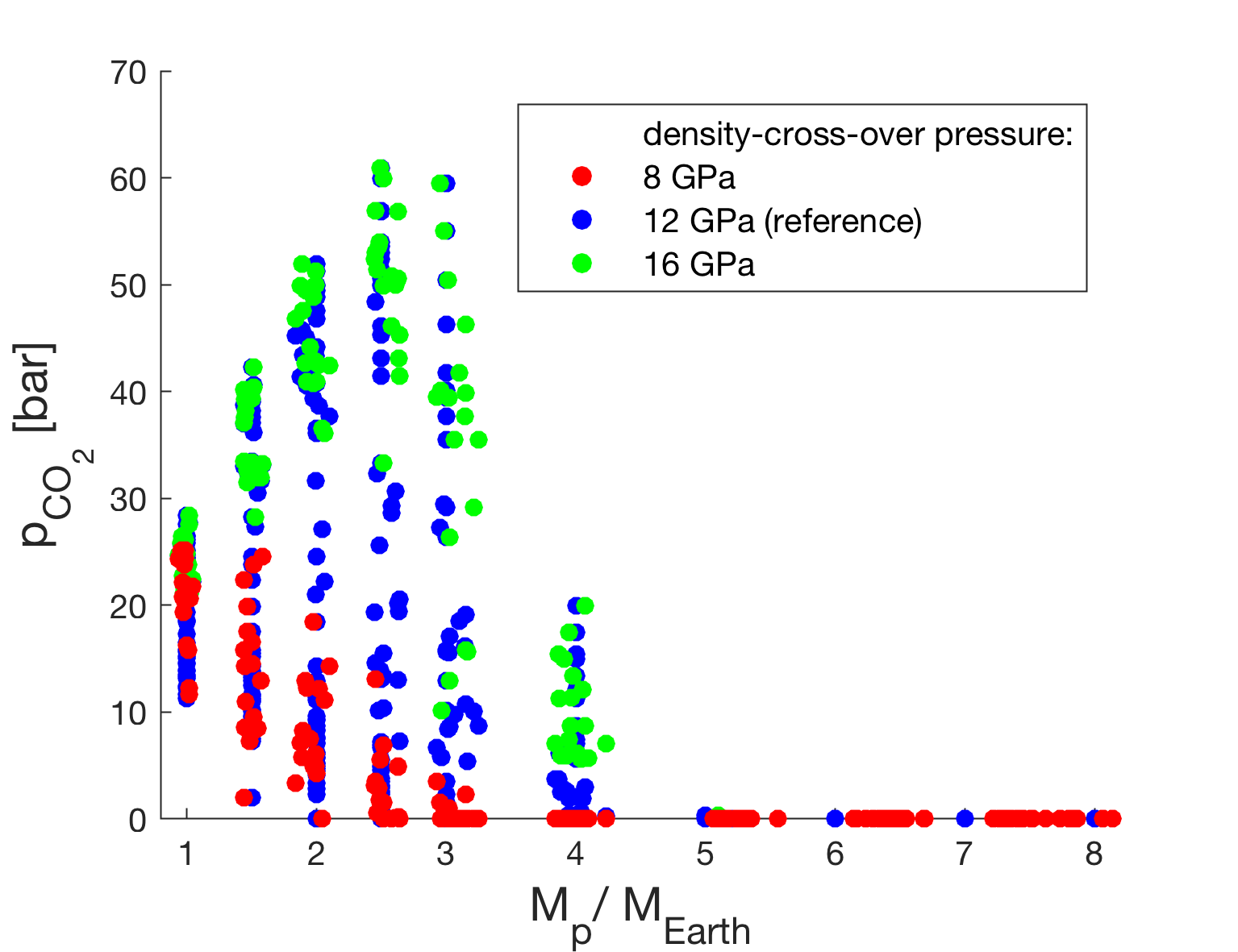}
   \caption{ {Influence of density-cross-over pressure $P_{\rm cross-over}$ (red, green, and blue dots) on the amount of outgassing as a function of planet mass. Reference case and cases 12 and 13 are shown. }}
              \label{figCROSS}
    \end{figure}

 {When melt occurs, its density contrast to the residue determines whether the melt migrates to the surface where it outgasses. The pressure up to which melt rises due to its buoyancy is parameterized by the density-cross-over pressure $P_{\rm cross-over}$. Here we investigated the influence of $P_{\rm cross-over}$ on the amount of outgassing. Reasonable ranges for $P_{\rm cross-over}$ for anticipated variabilities of exoplanet mantle compositions are poorly understood. We test $P_{\rm cross-over}$ being equal to 8, 12 (reference), and 16 GPa, inspired by theoretical and empirical studies \citep{sakamaki2006stability,bajgain2015structure}.
The resulting effect on the amount of outgassed $\rm CO_2$ is shown in Figure \ref{figCROSS}. As expected, smaller values of $P_{\rm cross-over}$ will lead to a reduced region where buoyant melt can exist and thus reduces outgassing (and vice versa). 

At small planet masses (<2 \ME), where the depletion is most effective, an increase of $P_{\rm cross-over}$  has marginal effects on outgassing. A significant influence on $p_{\rm CO_2}$ is only seen for the lower limit of $P_{\rm cross-over} = 8$ GPa.
At {intermediate} planet masses (2-4 \ME), where outgassing is dominantly pressure-limited (see Section \ref{versusmass}), the density-cross-over pressure can significantly alter the amount of outgassed $\rm CO_2$. For high planet masses above 5 \ME, volcanic outgassing is not effected by $P_{\rm cross-over}$.
}

    \subsection{Outgassing versus hydration or mantle rock}
    
\begin{figure}
   \centering
   \includegraphics[width = .5\textwidth, trim = 0cm 0cm 0cm 0cm, clip]{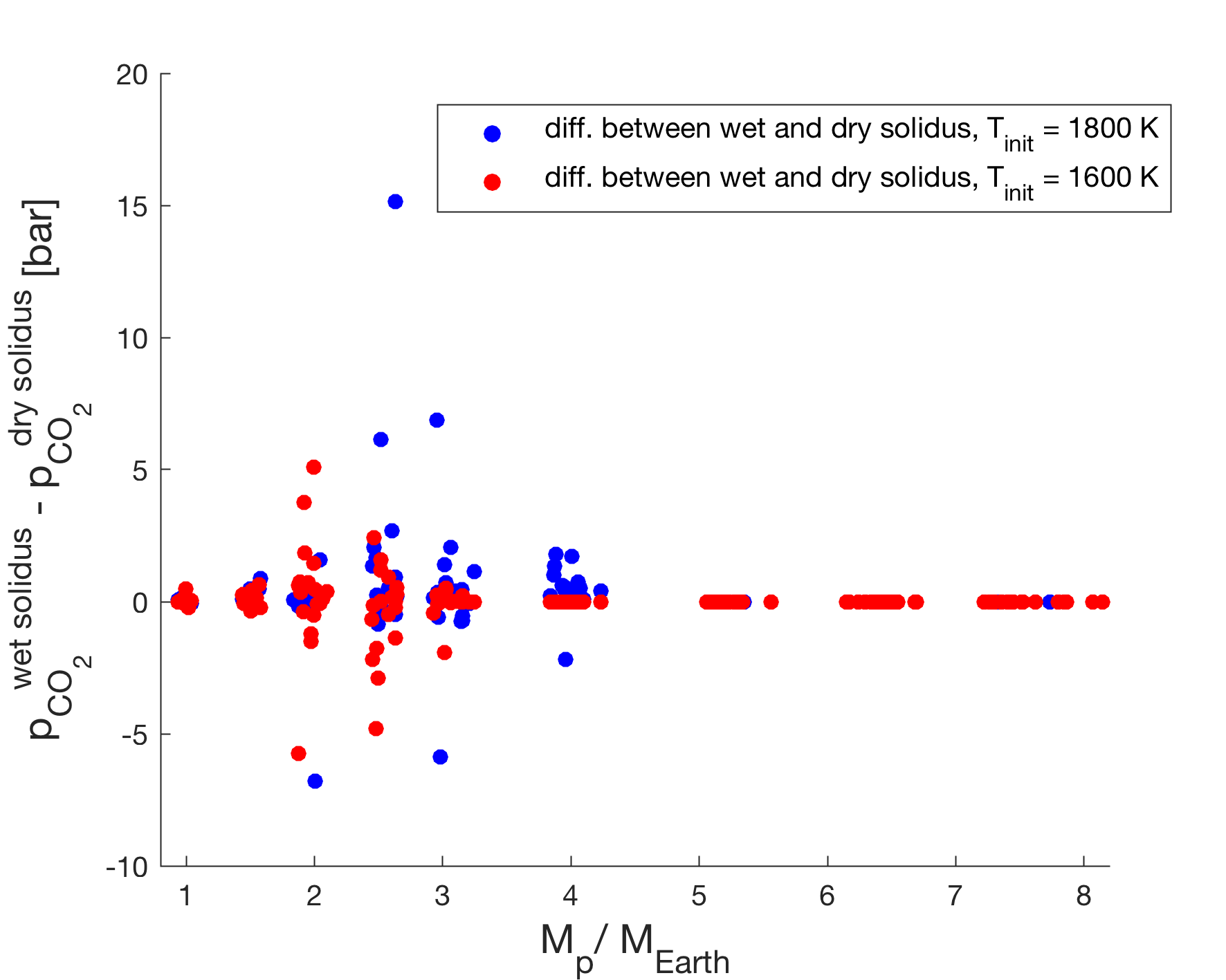}
   \caption{ {Influence of rock hydration on outgassing: the difference in the amounts of outgassed $\rm CO_2$  between  dry and  wet solidus melting temperatures are plotted versus planet mass for two different initial mantle temperatures $T_{\rm init}$ (1600 K and 1800 K). The shown differences are comparisons between case 10 and the reference case (blue dots) as well as case 11 and 4 (red dots).}}
              \label{figWET}
    \end{figure}

 {Little influence on volcanic outgassing is seen by accounting for hydration of rocks as illustrated in Figure \ref{figWET}. 
{We investigate the influence of a hydrated mantle (leading to a reduced solidus melting temperature, see Section \ref{convection}) for two different initial mantle temperature profiles (cases 10 and 11). }
Water partitions very easily into the melt already for small fractions of partial melting. 
This results in rock being dehydrated very quickly and water being extracted during the early evolution. Therefore, the resulting amount of outgassed $\rm CO_2$ is only weakly effected by hydration of rocks.
Over their lifetime, the amount of outgassed $\rm CO_2$ departs by less than 5 bars due to rock hydration for the majority of super-Earths.}

\subsection{Resolution}

   \begin{figure}
   \centering
   \includegraphics[width=1\hsize]{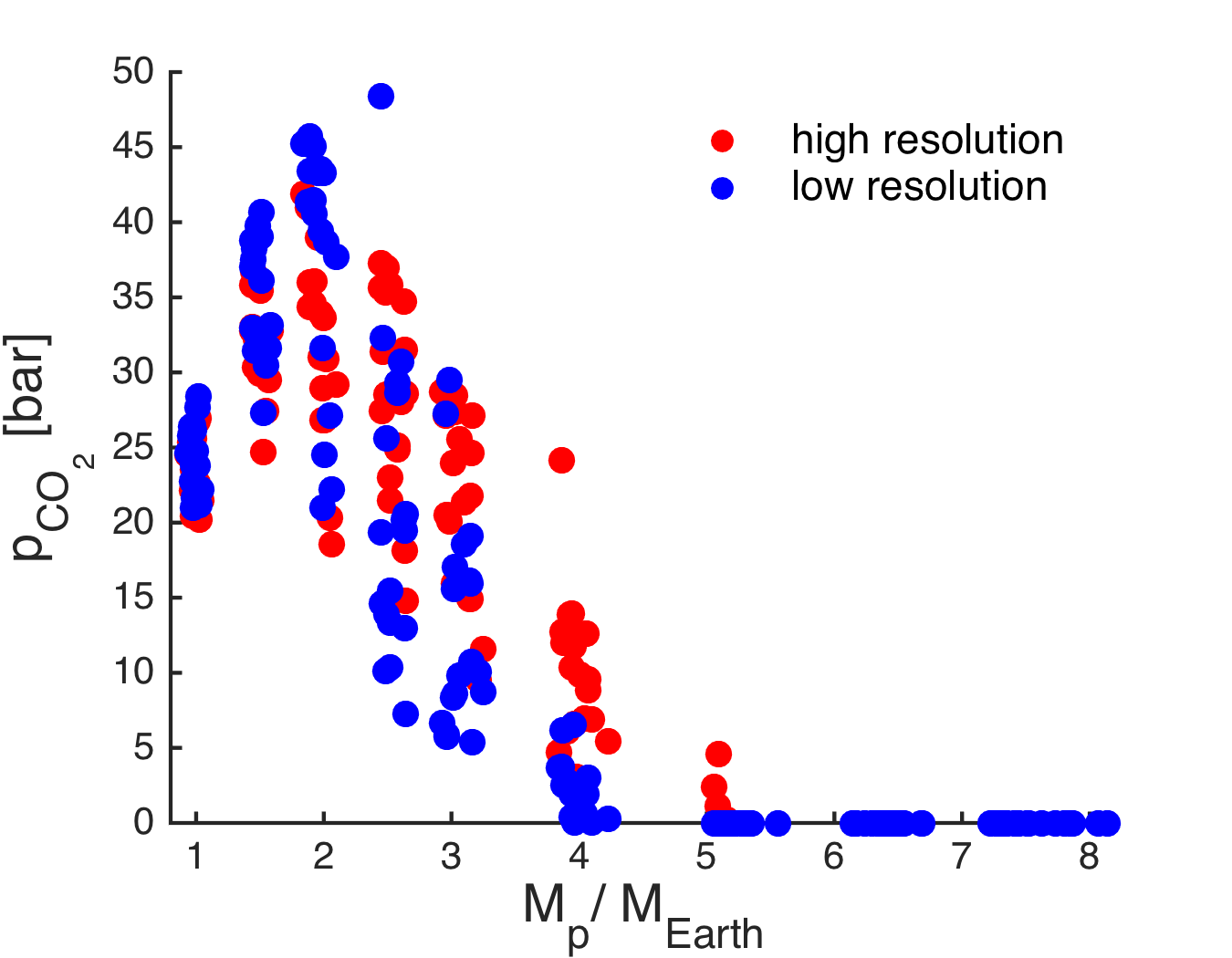}
   \caption{Influence of model resolution on simulated outgassing. The amount of outgassing of $\rm CO_2$ is denoted in terms of partial pressure $p_{\rm CO_2}$. The case 8 and the reference case are shown (Table \ref{tab:refCase}). Low-resolution (blue) refers to a radial resolution of 25 km, whereas high-resolution (red) refers to a 10 km resolution.}
              \label{fig:res}
    \end{figure}
    
The radial resolution in the convection model is fixed to 25 km in the reference case and is set to a higher resolution of 10 km in case 8 (see Table \ref{tab:refCase}). For small planet masses ($\leq$ 2 \ME) with extended melting regions, the higher resolution marginally effects the outcome. However, for higher mass planets (2--5\ME) a higher resolution allows to better capture the extend of melting zones and thus outgassing estimates are in average $\sim 18 \%$ (6 bar) higher. 

\section{Scaling of outgassing}

 {At an earlier stage of our study, we tried to describe the simulated outgassing using boundary layer theory only, however, we realized that the outgassing strongly depends on the internal temperature of the upper mantle, which is particularly poorly predicted by boundary layer theory. }
Here, we develop an empirical scaling law that  {uses boundary layer theory in parts} to predict the above studied trends of mantle depletion and outgassing based on the large number of simulations. We focus on parameters that have first and second order effects on depletion and outgassing.  Our proposed functional form for a scaling is underpinned by the following physical relationships, in which we introduce scaling parameters (i.e., $\alpha, \beta, \gamma,  \zeta, \zeta_1, \zeta_2, \zeta_3, \zeta_4, \zeta_5, \nu, \theta, \kappa, \lambda, \mu, \xi, \psi, \omega$).

Since melt depletion occurs at pressures below the cross-over pressure ($P_{\rm cross-over}$) and temperatures above solidus temperatures $T_s$, we consider mantle depletion to be
\begin{equation} \label{eq:d}
d =  \beta \cdot (V_{\rm Pcross}/V_{\rm mantle})^\alpha \cdot(T_{\rm eff} - T_{\rm s})^\lambda ,
\end{equation} 
where $V_{\rm Pcross}$ is non-negative and is the part of the total mantle volume $V_{mantle}$ which is below the lid and in which pressures are below $P_{\rm cross-over}$,
\begin{equation} \label{eq:Vcross}
V_{\rm Pcross} \approx 4/3 \cdot \pi \cdot \left((R_{\rm p}-  {\delta_{\rm LAB}})^3 - \left(R_{\rm p} - \frac{P_{\rm cross-over}}{{\rm d}p/{\rm d}z}\right)^3 \right) ,
\end{equation} 
where ${\rm d}p/{\rm d}z$ is the pressure gradient.  {The depth $\delta_{\rm LAB}$ of the boundary between the rigid lithosphere and the ductile asthenosphere depends on viscosity $\eta$ and is approximated \rev{using the asymptotic solutions of the Stokes equation \citep{Reese98}}:}
\begin{equation} \label{eq:delta}
 {
\delta_{\rm LAB} = \psi  R_{\rm p} \Delta\eta^{0.2}.
}
\end{equation} 
 {
\rev{We obtain a best-fit value for the exponent of 0.2 in Equation \ref{eq:delta}, which is similar to the exponent derived for Newtonian convection from asymptotic boundary layer theory \citep{fowler85,solomatov95,Reese98} and numerical studies \citep{reese1999stagnant}. The value is below the classical exponent of 1/3 derived for steady-state boundary layer theory \citep{solomatov95}, since our simulations are time-dependent and use a temperature-and pressure-dependent viscosity \citep{huttig2011regime}.}  Combining equation \ref{eq:Vcross} and \ref{eq:delta}, we obtain
\begin{equation} \label{eq:Vcross2}
\begin{split}
V_{\rm Pcross} \approx \mbox{ }& 4/3 \cdot \pi \cdot \Bigg(R_{\rm p}^3(1- \psi\Delta\eta^{0.2})^3 - \\
&\left(R_{\rm p} - \frac{P_{\rm cross-over}}{{\rm d}p/{\rm d}z}\right)^3 \Bigg) ,
\end{split}
\end{equation} 
}
The volume of the mantle is defined by
\begin{equation} 
V_{\rm mantle} = 4/3 \cdot \pi \cdot \big(R_{\rm p}^3 - ~\rcM^3 \big).
\end{equation}

The solidus temperature varies depending on the iron mass fraction, which is discussed in Section \ref{convection}. 
\begin{equation} \label{eq:ts}
T_{\rm s} = \zeta_1 + \Delta_{T_{\rm s}} = \zeta_1 + 360\cdot {(0.1-\mathcal{X}_{\rm Fe})}.
\end{equation}
The effective mantle temperature $T_{\rm eff}$ represents a time-averaged temperature, which we intend to use for the scaling.
We assume that $T_{\rm eff}$ depart linearly from a reference case depending on both the initial mantle temperature $T_{\rm init,mantle}$ and the amount of radiogenic heating sources, with $T_{\rm init,mantle}=1800$K and $Q_{\rm rad}=1 c_E$ being reference values.
\begin{equation} \label{eq:teff}
\begin{split}
T_{\rm eff} = \mbox{ }& \zeta_2 + \zeta_3 {(T_{\rm init,mantle}-1800 {\rm K})} + \\ &\zeta_4 {(Q_{\rm rad}-1 c_{\rm E})} + \zeta_5~\log(\Delta_{\eta}).
\end{split}
\end{equation}

Also, we account for the influence of $\mgsi$ and core size \rc on depletion, as discussed in section \ref{interiori}. A linear influence of $\mgsi$ on depletion is appropriate given our test models. We find that the influence of core size (\rc) and mantle iron content ($\mathcal{X}_{Fe}$) can better predict depletion (by 20\%), when a second order term is used that involves radiogenic heating sources. This is because for low radiogenic heating (case 3 in Table \ref{tab:refCase}), we do not observe significant influences of \rc and $\mathcal{X}_{Fe}$ on depletion.

\begin{figure}
   \centering
   \includegraphics[width=1.0\hsize, trim = 0.2cm 0cm 1.4cm 0cm, clip]{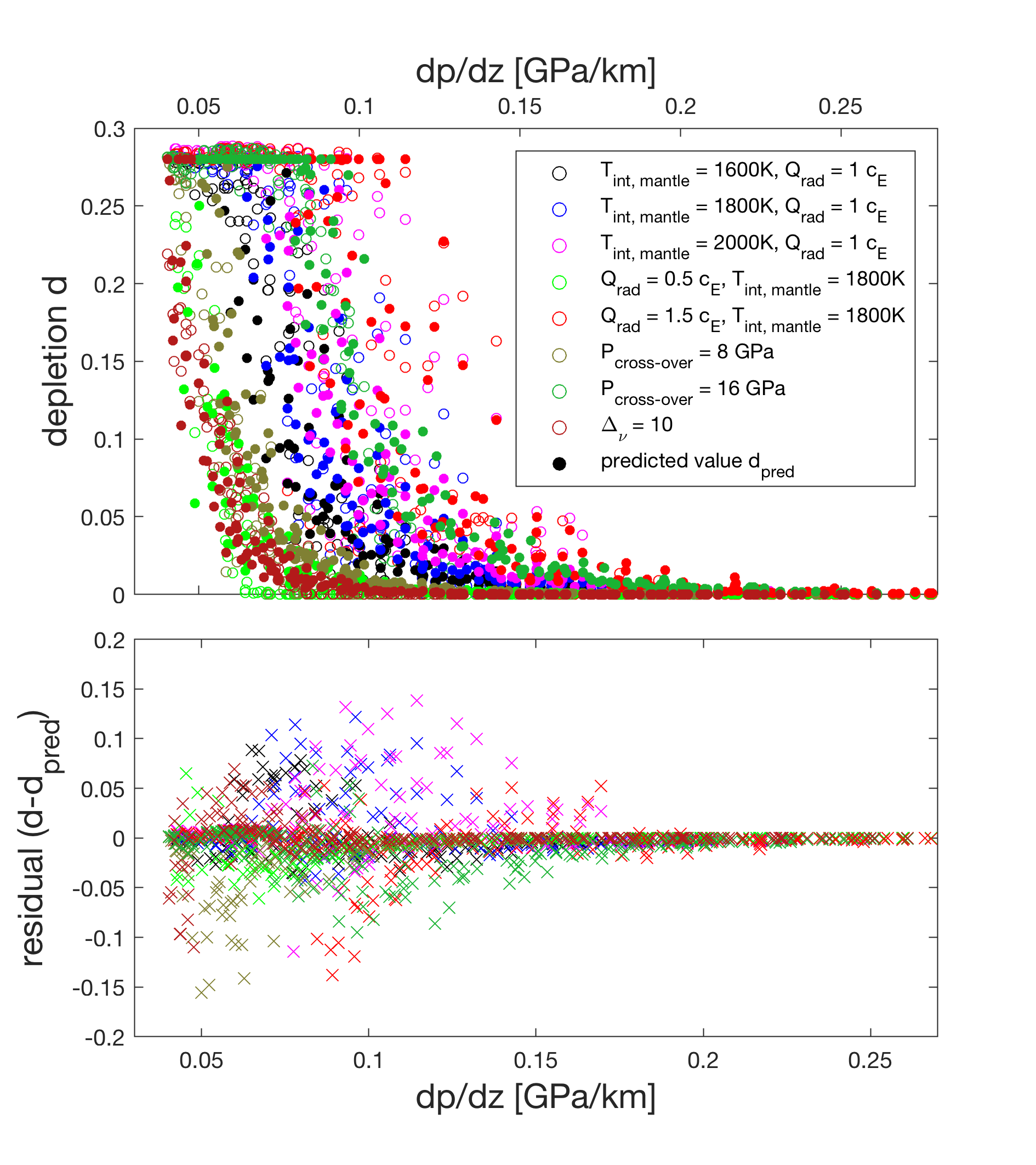}
   \caption{Fit between simulated and predicted mantle depletion using equation \ref{eq:fitd2} (upper panel) and corresponding residuals (lower panel). }
              \label{fig:deplfit}
    \end{figure}

On this basis and by combining the above equations as well as normalizing the linear scaling factors by the reference values, we finally obtain:
 {
\begin{equation} \label{eq:fitd2}
\begin{split}
d_{\rm pred} = \mbox{ }& {\rm max}\Bigg\{0,{\rm min}\bigg\{d_{\rm max},  (V_{\rm Pcross}(\psi)/V_{\rm mantle})^\alpha \cdot \beta \\
 & \cdot \bigg(1 + \gamma \frac{(T_{\rm init,mantle}-1800 {\rm K})}{1800 {\rm K}}  \\
 & + \zeta \frac{(Q_{\rm rad}-c_{\rm E})}{c_{\rm E}} \\
 & + \omega \log(\Delta\eta)\\
  & +\nu \frac{{\rm Mg/Si}_{\rm bulk}}{{\rm Mg/Si}_{\rm Sun}} \\
 & -\theta\cdot\frac{ (\mathcal{X}_{\rm Fe}-0.1)}{0.1} \cdot \frac{(Q_{\rm rad}-\mu \cdot c_{\rm E})}{(c_{\rm E}- \mu\cdot  c_{\rm E})}\\
 & +\kappa \frac{r_{\rm core}}{R_{\rm p}}\cdot \frac{(Q_{\rm rad}-\mu\cdot c_{\rm E})}{(c_{\rm E}- \mu \cdot c_{\rm E})}\bigg)^\lambda \bigg\}\Bigg\} \,.
\end{split}
\end{equation} 
}
 {where $V_{\rm Pcross}(\psi)$ refers to equation \ref{eq:Vcross2}}.
The scaling parameters (i.e., $\alpha, \beta, \gamma,$ etc.) and their fitted values are listed in Table \ref{tab:estimate}.
We use a nonlinear regression model (i.e., the {\it fitnlm} function of MATLAB) in order to determine the scaling parameters such that the root mean squared error (L2-norm) of the difference between simulated $d$ and $d_{\rm pred}$ is minimized.  We find a root-mean-square (RMS) error of  {0.028}. 
The fit and associated residuals between simulated and predicted depletion is depicted in Figure \ref{fig:deplfit}. The quality of the fit is limited due to the statistical nature of the interior model selection that results in moderate scatter which equation \ref{eq:fitd2} does not fully capture.

\begin{figure}
   \centering
   \includegraphics[width=1\hsize, trim = 0.2cm 0cm 1.4cm 0cm, clip]{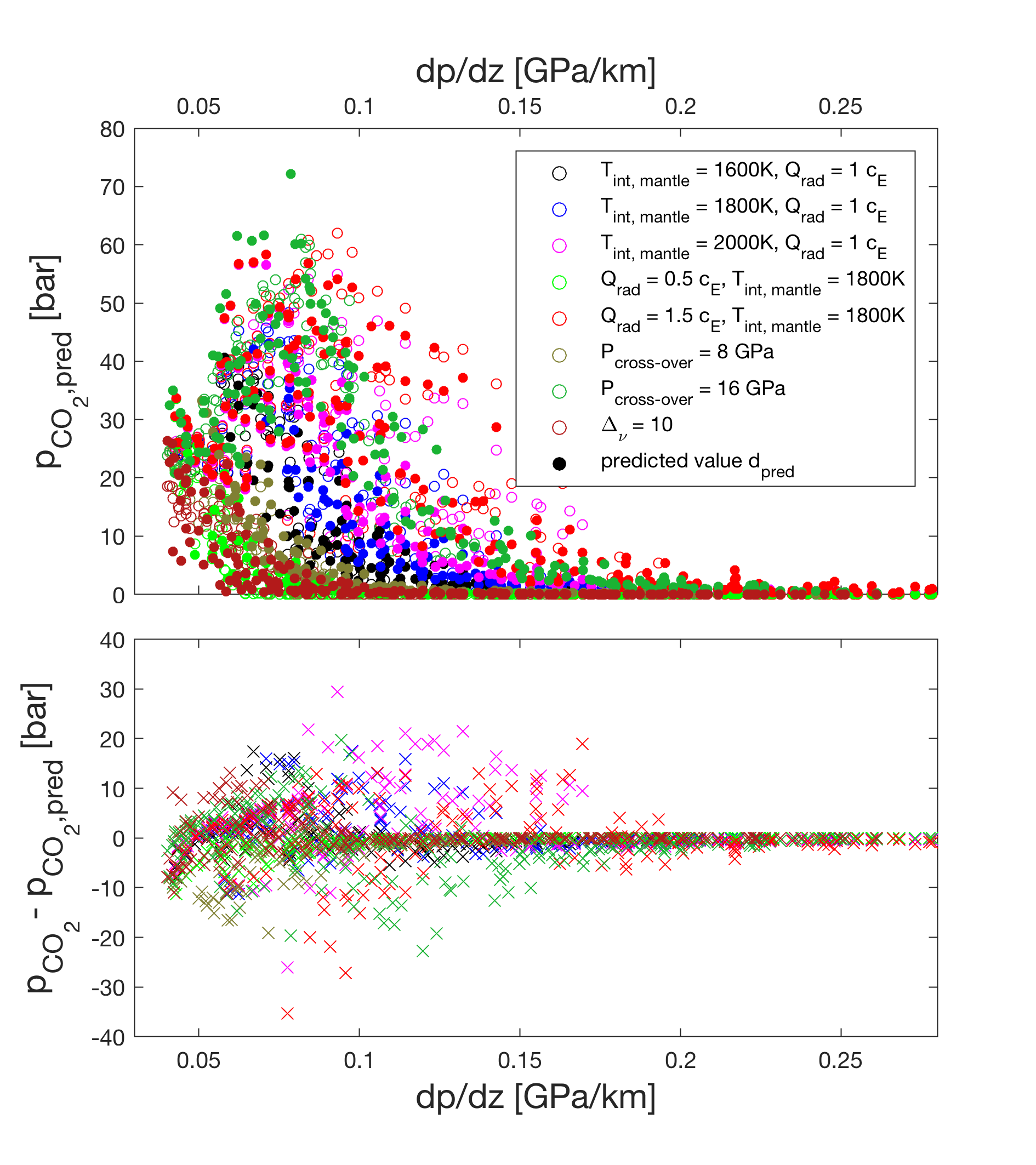}
   \caption{Fit between simulated and predicted amounts of outgassed CO$_2$ using equation \ref{eq:fitd2} and \ref{eq:pco2} (upper panel) and the corresponding residuals (lower panel).}
              \label{fig:outgfit}
    \end{figure}

The amount of total outgassed CO$_2$ in terms of partial pressure ($p_{\rm CO_2}$) is proportional to the mass of outgassed CO$_2$. The mass of CO$_2$ depends on depletion $d$ and the amounts of outgassed volatiles in the mantle, which is constant in all cases (here: 1000 ppm, see Table \ref{tab:refCase}).
\begin{equation} \label{eq:mco2}
m_{\rm CO_2} \propto d \cdot V_{\rm mantle},
\end{equation} 
and using a restatement of Newton's second law the  pressure that corresponds to the mass of outgassed CO$_2$ is described as
\begin{equation} 
p_{\rm CO_2} = \frac{m_{\rm CO_2} \cdot g} { 4 \pi R_{\rm p}^2 },
\end{equation} 
and thus the predicted $p_{\rm CO_2}$ in bar can be written as
\begin{equation} \label{eq:pco2}
p_{\rm CO_2,pred} =  \xi \cdot  10^{-5} \cdot \frac{d_{\rm pred} \big(R_{\rm p}^3 - ~\rcM^3\big) G M_{\rm p}}{ 3  R_{\rm p}^4 }.
\end{equation} 
where $\xi$ is another scaling parameter, $G$ is the gravitational constant, and the factor $10^{-5}$ accounts for the conversion from SI-units to bar.
While using the predicted mantle depletion $d_{\rm pred}$ in the above equation, we do another nonlinear regression to determine $\xi$ in order to best fit $p_{\rm CO_2}$ by $p_{\rm CO_2,pred}$. We expect $\xi$ to be on the order of the multiplication of $f_{\rm ex} \cdot f_{\rm CO_2} \cdot \bar{\rho}_{mantle}$ (see Table \ref{tab:refCase}), which is approximately 0.5 for a mean mantle density $\bar{\rho}_{mantle}$ of 5000 kg/m$^3$. Indeed, our estimate for $\xi$ of  {0.786} is on the same order. Figure \ref{fig:outgfit} illustrates the quality of the fit for $p_{\rm CO_2,pred}$, which is mostly limited by the residual scatter in $d_{\rm pred}$ and has a RMS of  {5.43}. 

In addition to the proposed scaling laws, we extensively tried different functional forms, including non-linear formulations and second order linear combinations of all parameters, but did not obtain significantly better fits.

 \begin{table}[h]
\caption{Estimates of scaling parameter. The standard deviation is denoted with $\sigma$, and the quality of fit with the $p$-value. Note that the significance of parameter estimates are only marginal in case of $p$-values larger than 0.05. \label{tab:estimate}}
\begin{center}
\begin{tabular}{llll} 
\hline\noalign{\smallskip}
parameter & estimate & $\sigma$ & $p$-value\\
\noalign{\smallskip}
\hline\noalign{\smallskip}
\multicolumn{4}{c}{(10 parameters, Eq. \ref{eq:fitd2} \& \ref{eq:pco2}):}\\
\hline\noalign{\smallskip}    
 $\alpha$ &4.868 & 0.045 & 0\\ 
  $\beta$ &45032.612 & 8$\times10^{-7}$ & 0\\ 
  $\gamma$ &2.50 & 0.20 & 2$\times10^{-34}$\\ 
$\zeta$ & 0.843 & 0.08 & 1$\times10^{-27}$\\ 
  $\nu$& 0.038 & 0.01 & 3$\times10^{-7}$\\ 
 $\theta$&0.164 & 0.02 & 6$\times10^{-16}$\\ 
 $\kappa$ &0.486 & 0.06 & 2$\times10^{-17}$\\ 
$\lambda$ &2.968 & 0.17 & 3$\times10^{-62}$\\ 
$\mu$&0.721 & 0.01 & 0\\ 
$\xi$&0.786 & 0.02 & 5$\times10^{-250}$\\ 
 { $\psi$}&-0.002 & 0.0005 & 0.00021\\ 
 {  $\omega$}&-0.410 & 0.02 & 7$\times10^{-69}$\\ 
\hline\noalign{\smallskip}
\end{tabular} 
\end{center}
\end{table}

   \begin{figure}
   \centering
   \includegraphics[width = .5\textwidth, trim = 0cm 0cm 0cm 0cm, clip]{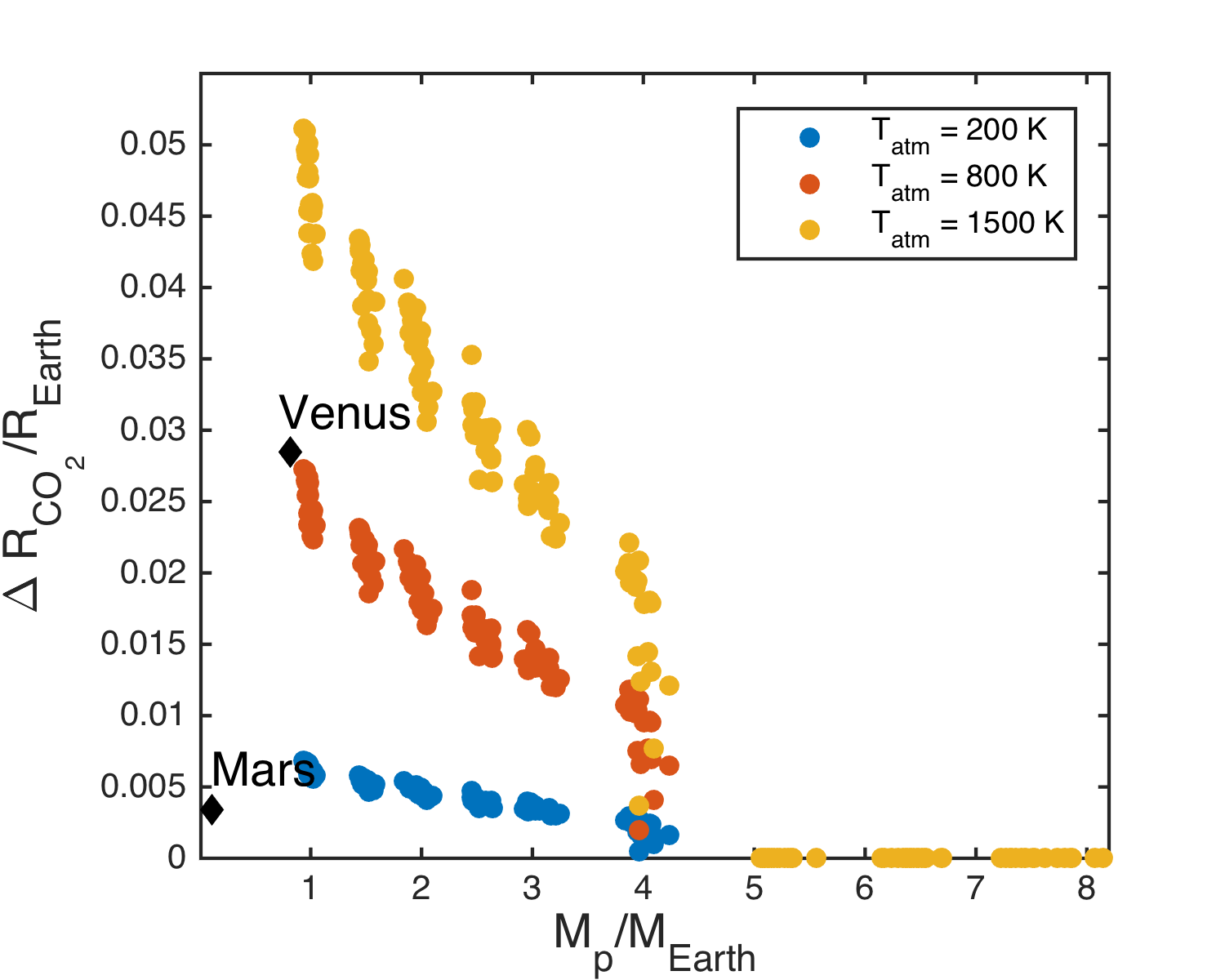}
   \caption{Gas layer thicknesses corresponding to calculated partial pressures $p_{\rm CO_2}$ for the reference case (Table \ref{tab:refCase}) assuming different atmospheric mean temperatures of 300 K, 800 K and 1500 K. Venus and Mars are shown for reference (Venus: $H$ = 15.9 km, $p_{\rm CO_2} = 92$~bar, $p_{\rm min}= 1$~mbar, $T_{\rm atm}= 737$~K; Mars: $H$ = 11.1 km, $p_{\rm CO_2} = 6.9$~mbar, $p_{\rm min}= 1$~mbar, $T_{\rm atm}= 210$~K).}
              \label{figDelta}
    \end{figure}

    \paragraph{Gas layer thicknesses}
\label{thickgas}
We demonstrated that the amount of outgassing is most efficient around $\sim$ 2~\ME, where highest values of  $p_{\rm CO_2}$ can be observed. In Figure \ref{figDelta}, we demonstrate how the distribution of $p_{\rm CO_2}$ would translate to gas layer thicknesses. We calculate the thicknesses $\Delta R_{\rm CO_2}$ using a scale height model similar to the model in \citet{dorn2017generalized}: 
\begin{equation}
\Delta R_{\rm CO_2} = H \ln{\left(\frac{p_{\rm CO_2}}{p_{\rm min}}\right)},
\end{equation}
where $p_{\rm min}$ is the pressure at which the atmosphere becomes opaque, that we simply fix to 1 mbar. The pressure scale height $H$ is calculated assuming a CO$_2$ atmosphere (mean molecular weight of 44.01 g/mol) and using a mean atmospheric temperature $T_{\rm atm}$,
\begin{equation}
H = \frac{T_{\rm atm} R^{*}}{g \cdot 44.01~{\rm g/mol}},
\end{equation}
where $g$ is surface gravity and $R^{*}$ is the universal gas constant (8.3144598 J mol$^{-1}$ K$^{-1}$).

While $p_{\rm CO_2}$ first increases and then decreases with planet mass, the corresponding thicknesses $\Delta R_{\rm CO_2}$ always decrease with planet mass $M_{\rm p}$. This is because the scale height $H$ is inversely proportional to $g$, and thus $H \sim 1/M_{\rm p}$.
Our approximation of $\Delta R_{\rm CO_2}$ represents the thickness of the outgassed atmosphere, neglecting any primary or primordial atmosphere, chemical weathering,  {or atmospheric escape}. We compare $\Delta R_{\rm CO_2}$ with independent atmospheric estimates for Venus and Mars and find  good agreement (Figure \ref{figDelta}). Compared to our scaled estimates, the thicker atmosphere on Venus can be explained by catastrophic outgassing events, whereas the thinner atmosphere on Mars by atmospheric erosion.
\rev{We note that for both Venus and Mars, regassing of CO$_2$ into the mantle is precluded, which is also due to the lack of liquid surface water and plate tectonics.}
    
    
    \section{ {Time-dependency}}
\label{Atimedep}

   \begin{figure}
   \centering
   \includegraphics[width = .5\textwidth, trim = 0cm 0cm 0cm 0cm, clip]{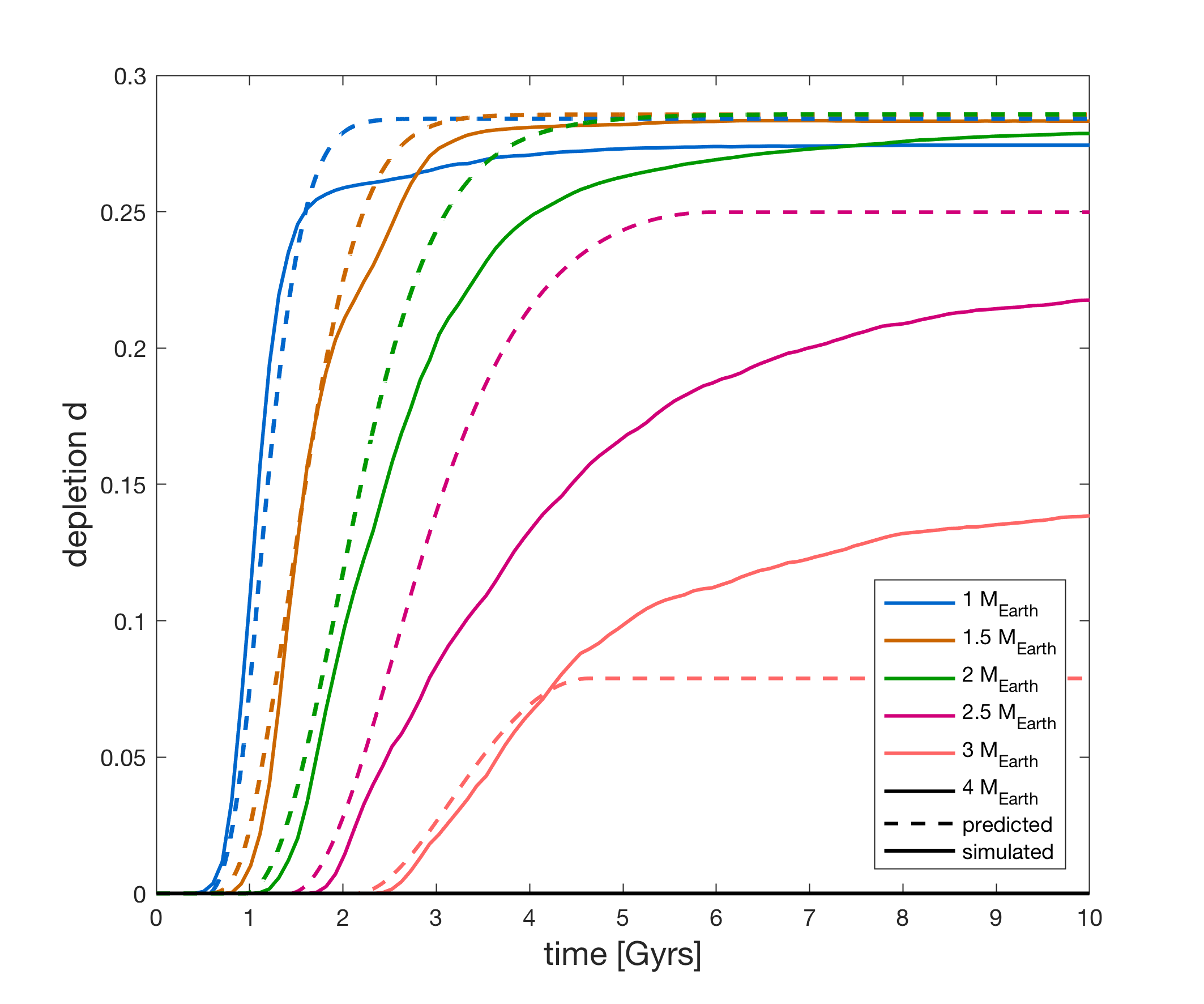}
   \caption{Time dependence of mantle depletion for selected planet masses (reference case). Solid lines show simulated mantle depletion, whereas dashed lines show predicted values based on our time-dependent scaling law.}
              \label{figtime}
    \end{figure}
    
 {Our empirical scaling law for depletion in Equation \ref{eq:fitd2} is not time-dependent. Yet, figure \ref{figtime} shows that we were able to nicely reproduce the time-dependence of depletion of our reference cases using a simple model based on boundary layer theory which we describe in the following.}

 {In order to reproduce the temporal evolution of the depletion, we consider that the mantle is divided in three layers: the lithosphere, the CO$_2$ producing region (from the bottom of the lithosphere to the cross-over depth) and the non-depleting mantle (everything below the cross-over depth). We consider that the lithosphere thickness is governed by the Rayleigh number (Ra), as indicated by boundary layer theory \citep{fowler85,solomatov95,Reese98}. Since both the viscosity of the mantle and the melt fraction in the CO$_2$ producing region are temperature-dependent, we carefully model the evolution of a reference temperature throughout time.}

\subsection{ {Evolution of the temperature}}

 {We numerically integrate the evolution of the temperature of the CO$_2$ producing region using the simple form:
\begin{equation}
\label{eq:T}
    T(t) = T_0 + \int_{t'=0}^{t} \frac{\partial T}{\partial t'} dt'.
\end{equation}
Temperature only evolves as a function of radiogenic heating and cooling from the lithosphere:
\begin{equation}
\label{eq:dTdt}
    \frac{\partial T}{\partial t} = \frac{Q}{C_p} -\frac{\Phi S}{\rho C_p V_m},
\end{equation}
where $Q=Q_0\exp(-t/t_{1/2})$ is the radiogenic heating ($Q_0=2.42\cdot 10^{-11}$ W$\cdot$kg$^{-1}$ and the half life $t_{1/2}=2.85$ Gyr), $C_p$ is the heat capacity ($C_p=1200$ J$\cdot$kg$^{-1}\cdot$K$^{-1}$), $\Phi$ is the (time-dependent) heat flux, $S$ is the surface of the planet, $\rho$ is the average density of the planet and $V_m$ is the volume of the mantle.
}

 {The heat flux is computed using the standard boundary layer theory:
\begin{equation}
\label{eq:phi1}
    \Phi \propto \Phi_{\textrm{diff}} \textrm{Ra}^{n},
\end{equation}
where $n=0.28$ is consistent with previous studies \citep{fowler85,solomatov95,Reese98} $\Phi_{\textrm{diff}}$ is the diffusive heat flux at the surface (in the absence of convection):
\begin{equation}
\label{eq:phi2}
    \Phi_{\textrm{diff}} \propto \frac{R_\Earth}{R},
\end{equation}
where we considered a fixed "equilibrium" surface to core temperature difference for simplicity. To compute the Rayleigh number, we consider the planet mass-dependence of the average thermal expansivity, density, gravity and mantle thickness (assumed half of the planet radius):
\begin{eqnarray}
    R      & = &      R_\Earth \; M^{0.26}, \\
    \alpha & = & \alpha_\Earth \; M^{-1.43},\\
    \rho   & = &   \rho_\Earth \; M^{0.22}, \\
    g      & = &      g_\Earth \; M^{0.48},
\end{eqnarray}
where $M$ is the normalized planet mass $M=M_p/M_\Earth$. The scaling for the radius was previously derived in section \ref{versusmass}. Gravity $g$ was obtained using $g=GM_p/R^2$. The average density was estimated by dividing planet mass by planet volume (thus assuming that the compressibility of mantle and core are similar). The scaling for thermal expansivity $\alpha$ follows \citet{Katsura}: $\alpha \propto \alpha_\Earth  ({\rho_\Earth}/{\rho})^{\delta_T}$ with $\delta_T \approx 6$.
The viscosity below the lithosphere is approximated by:
\begin{equation}
\label{eq:eta}
    \eta = \eta_0 \exp\left(\frac{E}{R_b}\left(\frac{1}{T}-\frac{1}{T_0}\right)\right),
\end{equation}
where $\eta_0$ is a reference viscosity, $E$ is the activation energy ($E=300$ kJ/mol), $R_b$ is the universal gas constant and $T_0$ is a reference temperature ($T_0=1800$K).
The Rayleigh number can then be defined \citep{travis94}:
\begin{equation}
\label{eq:Ra}
    \textrm{Ra} = \frac{\alpha\rho g (R/2)^5Q}{\kappa \eta} \propto M^{0.57} \exp\left(-\frac{t}{t_{1/2}}-\frac{E}{R_b}\left(\frac{1}{T}-\frac{1}{T_0}\right)\right).
\end{equation}
Using Equations \ref{eq:phi1}, \ref{eq:phi2} and \ref{eq:Ra}, the heat flux becomes:
\begin{equation}
\label{eq:phi3}
\Phi = \Phi_0 M^{-0.1} \exp\left(-\frac{0.3t}{t_{1/2}}-\frac{0.3E}{R_b}\left(\frac{1}{T}-\frac{1}{T_0}\right)\right),
\end{equation}
where $\Phi_0$ is a constant. We found that a reference heat flux $\Phi_0=10$ mW$\cdot$m$^{-3}$ best fits the time-dependent formulation. Using Equation \ref{eq:T}, \ref{eq:dTdt} \ref{eq:phi3}, we were able to estimate the evolution of the temperature below the lithosphere for all planet masses.
}

\subsection{ {Evolution of the depletion}}

 {The depletion $d$ is considered to be the volume sum of depletions in the CO$_2$ producing region (top) and in the rest of the mantle (bot):
\begin{equation}
    d = \frac{d_{\textrm{top}}V_{\textrm{top}}+ d_{\textrm{bot}}(V_m-V_{\textrm{top}})}{V_m}
\end{equation}
where $V_m$ is the volume of the mantle and $V_{\textrm{top}}$ is the volume of the CO$_2$ producing region defined by:
\begin{equation}
    V_{\textrm{top}} = \frac{4\pi}{3}\left(\left(R-l\right)^3-R_{co}^3\right).
\end{equation}
$R_{co}$ is the cross-over radius above which melt becomes lighter than the solid and $l$ is the lithosphere thickness obtained using the heat flux:
\begin{equation}
    l = 0.78\frac{(T-300)k}{\Phi}, \label{eq:l}
\end{equation}
with $k$ the thermal conductivity ($k=3$ W$\cdot$m$^{-1}\cdot$K$^{-1}$). The factor 0.78 slightly diminishes the lithosphere thickness to account for radiogenic heating in the lithosphere and the potential topography of the based of the lithosphere. It was found necessary to slightly diminish the lithosphere thickness to obtain a consistent temperature evolution and volume of CO$_2$ producing region.
}

 {The depletion in the top region is obtained at each time $t$ in two stages. First, depletion $d_{\textrm{top}}$ is updated using the melt fraction $\phi$, itself derived from the temperature:
\begin{eqnarray}
    \phi & = & \frac{T-T_s}{T_l-T_s}, \label{eq:meltfrac}\\
    d_{\textrm{top}}(t) &=& \max\left(d_{\textrm{top}}(t-\Delta t),0.3\phi\right),
\end{eqnarray}
where $T_s$ is the solidus temperature (assumed to be 2100~K for simplicity) and $T_l$ is the liquidus temperature ($T_l=2300$K).
The melt fraction $\phi$ is kept between 0 and 1. The depletion of the previous time step is used as a minimum to prevent depletion to disappear if the melt fraction decreases.}

 {In each time step, the decrease of depletion is possible through the second stage in which exchange of mass between top and bottom mantle is explicitly estimated. Both top and bottom depletions are updated together using time substeps. An advective depletion flux is considered at the base of the CO$_2$ producing region. The evolutions of depletion take the form:
\begin{eqnarray}
    \frac{\partial d_{\textrm{bot}}}{\partial t} &=& 1.2\frac{V_{\textrm{top}}}{V_m-V_{\textrm{top}}}\frac{1}{R_p - r_{\rm core}}v\left(d_{\textrm{top}}-d_{\textrm{bot}}\right), \label{eq:ddbdt}\\
    \frac{\partial d_{\textrm{top}}}{\partial t} &=&-\frac{\partial d_{\textrm{bot}}}{\partial t} \frac{V_{\textrm{top}}}{V_m-V_{\textrm{top}}}\label{eq:ddtdt},
\end{eqnarray}
where $v$ is a velocity consistent with the heat flux as prescribed by classical boundary layer theory \citep{fowler85,solomatov95,Reese98}:
\begin{equation}
    v = v_0 \left(\frac{\Phi}{\Phi_0}\right)^2
\end{equation}
where $v_0$ is fixed to 1 cm/yr. Equation \ref{eq:ddbdt} shows that the propagation of depleted material from the CO$_2$ producing region to the rest of the mantle does not only depend on the velocity over the thickness of the mantle. The volume ratio of top over bottom layers has to be considered as well. Indeed, if the CO$_2$ producing region is very thin, only a thin layer of depleted material will propagate in the mantle. The evolution of the depletion in the top (Equation \ref{eq:ddbdt}) is equal to minus the evolution in the bottom multiplied by the volume ratio to conserve the mass of depleted material during advection.
}

 {Figure \ref{figtime} shows the resulting \rev{evolution of} depletion in which it has been considered that the top 100~km is fully depleted; depletion has therefore been multiplied by the volume of the mantle below 100~km depth over the total volume of the mantle.}
 
    \begin{figure}
   \centering
   \includegraphics[width = .5\textwidth, trim = 0cm 0cm 0cm 0cm, clip]{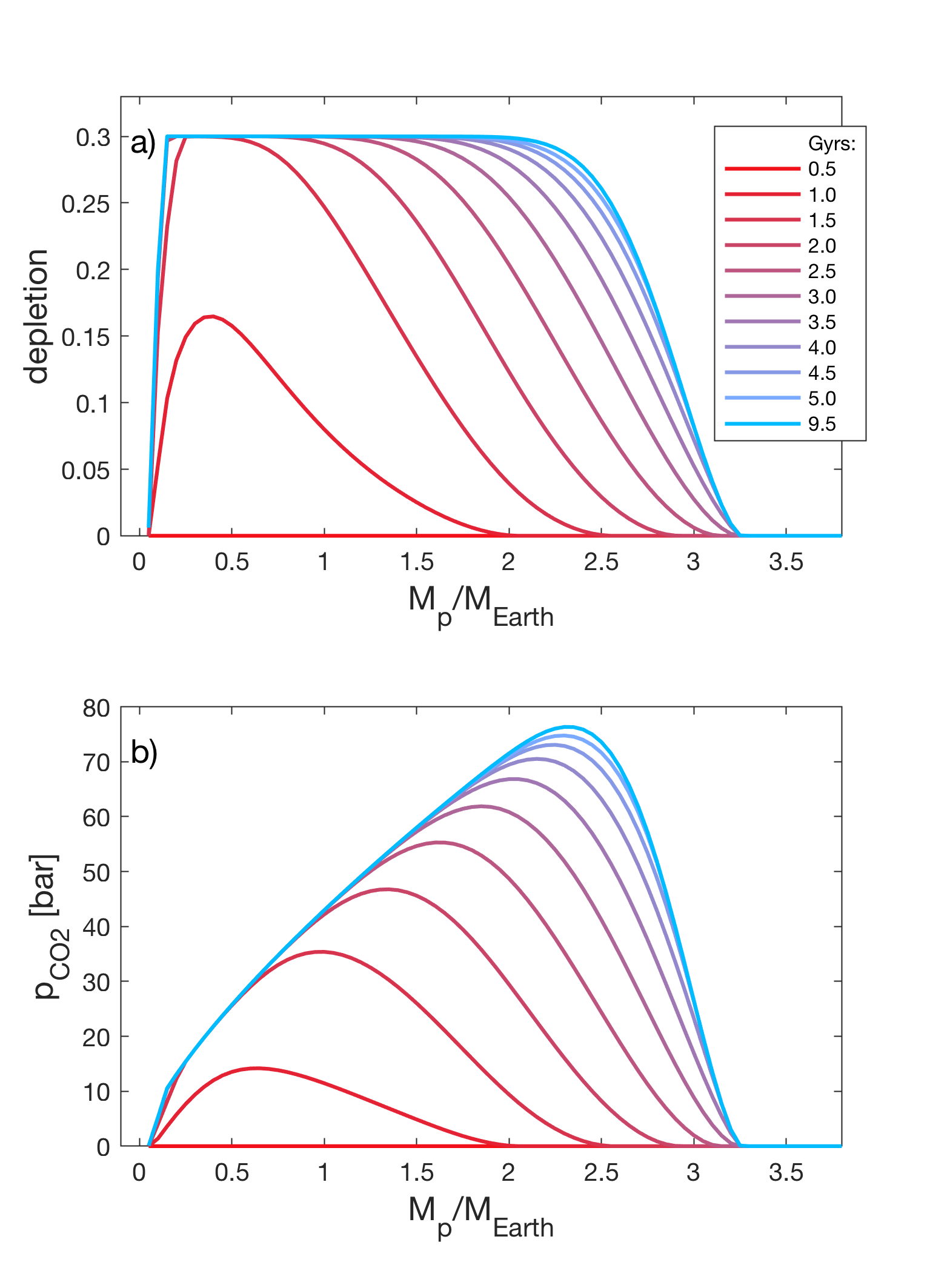}
   \caption{\rev{Time dependence of (a) mantle depletion and (b) amount of outgassed CO$_2$ as a function of planet mass (reference case). Depletion has been calculated with our analytical formulations in Section \ref{Atimedep}, and Equation \ref{eq:pco2} was subsequently used to calculate the amount of outgassed CO$_2$.}}
              \label{newTime}
    \end{figure}
    
 \rev{In Figure \ref{newTime}, we show the time-evolution of depletion and outgassing for various planet masses (of the reference case). The curves of depletion were calculated using our analytical formulation described above. The amount of outgassed CO$_2$ were then calculated using Equation \ref{eq:pco2}. For ages below 4.5 Gyrs, we see that planetary age has a first order influence on depletion and $p_{\rm CO_2}$. Variations beyond 4.5 Gyrs are small. This is important inasmuch observed exoplanets have a wide range of ages. }

\subsection{ {Comparison between the time-dependent depletion model and the scaling law for depletion for a fixed 4.5 Gyr evolution}}

 {Since the evolution of depletion relies on the numerical integration of partial differential equations, \rev{Equation \ref{eq:fitd2} does not provide an explicit time-dependent prediction for depletion}. Yet the necessity for most terms in Equation \ref{eq:fitd2} can be better understood considering the driving processes in our time-dependent formulation.}

 {The first term of equation \ref{eq:fitd2} shows that depletion is strongly related to the relative volumes of CO$_2$ producing region and the overall mantle. In our time-dependent formulation, these relative volumes also play a central role. We observed that the time-dependence of the existence of this top region is necessary to reproduce the onset times of depletion (see Figure \ref{figtime}). The existence of the top region strongly depends on both lithosphere thickness and internal temperature which are both strongly linked to the surface heat flux. The fact that the heat flux derived from boundary layer theory enables to reproduce the time-dependence of our simulations is remarkable and gives us confidence that boundary layer theory can be used in the investigation of exoplanet evolution. Yet, several terms of our scaling law for depletion (Equation \ref{eq:fitd2}) are very non-linear and cannot easily be derived from our time-dependent formulation which requires the use of convection simulations.}

 {The second term of Equation \ref{eq:fitd2} can also easily be understood using our time-dependent formulation. The occurence of melting and CO$_2$ degassing depends on whether or not the solidus temperature is reached. Equation \ref{eq:T} demonstrates the importance of the initial temperature to reach the solidus temperature $T_s$.}

 {The third term of Equation \ref{eq:fitd2} shows that internal heating also governs the occurence of melting. Again this can easily be understood from Equations \ref{eq:T}, \ref{eq:dTdt} and \ref{eq:meltfrac}.}

 {The importance of the viscosity (fourth term in Equation \ref{eq:fitd2}) in mantle depletion arises from several processes in our time-dependent formulation. The heat flux is related to the viscosity as demonstrated by boundary layer theory (Equations \ref{eq:phi1} and \ref{eq:Ra}). The heat flux plays a central role in both temperature evolution (Equation \ref{eq:dTdt}) and lithosphere thickness (Equation \ref{eq:l}). The negativity of the factor $\gamma$ in Equation \ref{eq:fitd2} shows that an increase in viscosity decreases the depletion. This shows the central role of existence of the CO$_2$ producing region as a large viscosity will result in a low heat flux and a large temperature. Melting is therefore more important (Equations \ref{eq:T}, \ref{eq:dTdt} and \ref{eq:meltfrac}) but as the lithosphere is too thick there is no volcanism and outgassing of CO$_2$.}

 {The Mg/Si ratio does not enter our time-dependent formulation. We therefore cannot reproduce the 5th term in Equation \ref{eq:fitd2}. The impact of the Mg/Si ratio could probably be understood investigating its effects on the Rayleigh number through density variations.}

 {The 6th and 7th term of our scaling law for depletion are very non-linear and can hardly be understood from our simple time-dependent formulation. These terms show that some combination of internal heating, iron content and core size have a second-order effect on depletion. The impact of iron content can be understood from our time-dependent formulation as the solidus temperature is strongly Fe-dependent as shown in section \ref{convection}.}

 {In conclusion, the time-dependence of CO$_2$-outgassing in stagnant-lid planets can be understood using boundary layer theory for any planetary mass. One limitation of our model is that crust production is neglected which makes it impossible to observe the recycling of basaltic material in the mantle. This could easily be taken into account in a parameterized model by estimating the amount of basalt produced and comparing its volume to the volume of the lithosphere. If the volume of basalt exceeds the volume of the lithosphere, then the depletion of the mantle should be decreased as enriched material should be dripping back in the mantle from the base of the lithosphere. This would simply result in adding a source term in the depletion in the CO$_2$ producing region. However, this limitation has no impact on the large planets which anyway seem to never be able to produce basaltic material.}

\section{Discussion}
\label{Discussion}

Interior dynamics and outgassing are linked to interior properties. The anticipated variability of super-Earth interior structures and compositions can be partly informed by commonly observed astrophysical data from exoplanets. These data include planetary mass and radius, and bulk abundances of rock-forming elements (i.e., Fe, Mg, Si).
In addition, we expect a wide variability on key thermal parameters that are very difficult to constrain by observations. 
On this basis, we compiled a set of super-Earths that incorporates the anticipated variability of structural, compositional, thermal parameters, \rev{and age} of the majority of super-Earths. This set excludes super-Mercuries, that are distinct from super-Earths by larger core-mantle ratios. It is yet unclear how frequent super-Mercuries are.
 {Thus, our study goes beyond a simple parameter study where only one parameter is altered at a time. Our test cases incorporate both knowledge and ignorance on rocky exoplanet interiors.}

 \rev{This study is a significant step towards interpreting astrophysical  observations of exoplanet atmospheres by geophysical interior models. 
Any interpretation of astrophysical observations of super-Earths atmospheres must be done in light of relevant formation and  evolution processes of atmospheres. Here, we have focused on long-term outgassing processes that shape the atmosphere of terrestrial-type stagnant-lid exoplanets but neglect  other processes such as (1) the early outgassing from a magma ocean, or (2) atmospheric erosion due to stellar irradiation, (3) weathering, or (4) any primordial hydrogen atmosphere. {Considering these complexities, our results represent first order estimates.} We briefly discuss these aspects in the following.

\paragraph{Outgassing from magma oceans}
Early outgassing from a magma ocean \citep[e.g.,][]{elkins2008linked,lebrun2013thermal} can be in principle incorporated by choosing initial non-zero values of $p_{\rm CO_2}$. It is possible that such initial amounts of primary atmospheres exceed the variations due to the long-term volcanism by up to several 100 bars \citep{lebrun2013thermal}.

\paragraph{Atmospheric escape}
Although atmospheric escape can efficiently erode hydrogen atmospheres, the erosion of CO$_2$ is much more inefficient, also because their high mean-molecular weights \citep{lopez2016}. In addition, the Super-Earths of interest are temperate planets for which stellar irradiation is limited. If erosion of outgassed atmospheres is significant, than it is the early outgassed atmosphere from a magma ocean that is mainly effected, since the stellar high energy-irradiation is strongest during the early evolution of a star.
\paragraph{Weathering}
Outgassed CO$_2$ can also be removed from the atmosphere via carbonate weathering. However, weathering requires sufficient supply of fresh, weatherable rock, which is limited for stagnant-lid regimes \citep{foley2016}. \citet{foley2018carbon} argue that weathering can significantly limit atmospheric CO$_2$ accounting for supply-limited weathering. However, given the possible variability in Super-Earth's compositions, the variability in carbonation efficiency of different erupted rocks requires further understanding.

Other sinks of CO$_2$ are water oceans, however the carbon ocean reservoir is small compared to the mantle reservoir \citep{sleep2001carbon}. The solubility of CO$_2$ in water is temperature dependent and increases with lower temperatures \citep[e.g.,][]{kitzmann2015unstable, pierrehumbert2010principles}. In principle, our predicted amounts of outgassed CO$_2$ can be used as input in climate models to investigate whether CO$_2$ would be present as gas in the atmosphere, as ice on the surface, or partially dissolved in a possible water ocean \citep[e.g.,][]{menou2015climate,abbot2012indication,tosi2017habitability}. For Earth-sized stagnant-lid planets,\citet{foley2018carbon} suggest that CO$_2$ budgets low enough to prevent runaway greenhouse and high enough to prevent global glaciation range from $10^{-2} - 1$ times the Earth's budget.

\paragraph{Primordial hydrogen atmosphere}
  Any primordial hydrogen-dominated atmospheres could in principle make the identification of outgassed atmospheres difficult. Fortunately, even if spectroscopic investigations of a Super-Earth's atmosphere \citep[e.g.,][]{bourrier2017, benneke2016, knutson2014hubble} are not available, considerations of atmospheric escape \citep{dorn2017submitted} can provide necessary constraints in addition to mass and radius to distinguish between hydrogen-dominated and enriched (e.g., outgassed) atmospheres. Thereby, the thickness or mass fraction of a gas envelope that is likely outgassed from the interior can be quantified \citep{dorn2017submitted} and misinterpretations due to the presence of a hydrogen-dominated envelope can be reduced.

\paragraph{Observational constraints on outgassed atmospheres}

 Characterizing interiors and atmospheres of exoplanets is a highly degenerate problem. However, it is possible to quantify probabilities of atmospheric properties (i.e., mass and radius fraction of an atmosphere and its enrichment in heavier molecules) as demonstrated by \citet{dorn2017submitted}. They determine that enriched (and possibly outgassed) atmospheres preferably occur on planets of small masses and high equilibrium temperatures. Their use of a generalized Bayesian inference analysis allowed them also to quantify the atmosphere thicknesses for a set of about 20 exoplanets. Interpreting such a distribution of possibly outgassed atmospheres requires geophysical interior models. Our study provides a significant part of the necessary tools to perform an informed interpretation. 

Improved estimates on the distributions of possibly outgassed atmospheres are expected to be possible by the data of upcoming missions (e.g., TESS, CHEOPS, JWST). These missions will not only significantly increase the number of exoplanet detections (e.g., TESS), but also provide better precision on the data that we use to characterize their interiors (e.g., CHEOPS) and make it possible to probe in details the atmospheres of some tens of Super-Earths (JWST).

If observations confirmed our predicted trend of CO$_2$ atmospheres with planet mass, this would suggest that the majority of Super-Earths are in a stagnant-lid regime. Deviating behaviours may be explained by dynamic regimes other than stagnant-lid, e.g., plate tectonics \citep{valencia2007inevitability,kite2009geodynamics,korenaga2010likelihood,van2011plate,noack2014can,o2007geological,lenardic2012notion,foley2012conditions} or atmospheres being dominated by the early outgassing during the cooling of a magma ocean \citep{hamano2013emergence}.

Commonly observed exoplanets orbit at close distances to their stars which involves much higher surface temperatures than our assumed 280 K fixed value. In fact, surface temperatures of observed exoplanets may allow for surface rocks to be molten. Analyzing outgassing under such temperature conditions would require the modelling of a magma ocean, crustal production and melt migration processes, which is outside of the scope of this paper. Here, we focused on temperate exoplanets, for which upcoming missions (e.g., TESS, CHEOPS, JWST) will provide data for interior characterization, for example from planets around M-dwarf stars (e.g., Trappist-1 system).

\paragraph{Impact on Habitability}

The classical definition of the  habitable zone assumes the availability of greenhouse gases  such as CO$_2$ \citep[e.g.,][]{kasting1993,kopparapu2014habitable}. The outer boundary of the habitable zone mostly depends on the amount of  CO$_2$, while the inner boundary of the habitable zone is characterized by both the amounts of CO$_2$ and H$_2$O \citep{tosi2017habitability}. Since volcanism maintained over geological time-scales is possible for stagnant-lid planets, it is suggested that these planets can be habitable \citep[e.g.,][]{noack17melting, tosi2017habitability,foley2018carbon}.

Our results show that volcanism is limited for stagnant-lid planets of masses larger than 5-7 \ME or older than 5 Gyrs. This suggests that volcanic activity suitable for habitability is restricted to small planets ($<$5-7 \ME) as well as planets younger than $\sim$ 5 Gyrs. This is in agreement with previous studies \citep{noack17melting,foley2018carbon}.

Habitability depends also on the presence of other greenhouse gases that affect the surface temperature. Changes in surface temperatures feed back on the deformability of the lithosphere \citep[e.g.,][]{bercovici2014plate} and thus on outgassing. Possible greenhouse gases other than CO$_2$ that can drive this thermal feedback include for example H$_2$O. The efficiency of these feedback mechanisms depend on atmospheric amounts of the gases and their recycling dynamics between mantle and atmosphere. Here, we focused on the outgassing of CO$_2$ only. 
However, the limitations in volcanic activity discussed in our study similarly affects the outgassing of  gases other than CO$_2$. For example, the solubility of H$_2$O in melt is much higher than for CO$_2$. Therefore, partial pressures of outgassed H$_2$O can be one order of magnitude smaller compared to CO$_2$, while evolution trends of outgassing are similar \citep{tosi2017habitability}.  
 }

 \paragraph{Scaling law}
We developed scaling laws to summarize the efficiency of mantle outgassing depending on several Super-Earth characteristics.
The functional form of our derived scaling laws is based on physics and involves free fitting parameters. Our scaling law is able to describe the two trends of outgassing as a function of planet mass: at low mass, the outgassing increases with mass, whereas it decreases at higher masses. We showed that thermal, structural, and compositional parameters can alter the transition between these two trends.
We expect that other parameters that we did not consider could similarly affect this transition, however, they would not influence the existence of both trends. For example, rheological variability due to different grain sizes, hydration, compositions, or melt fraction are neglected in the present study.

 \paragraph{Thermal convection model}
As commonly done, the investigated stagnant-lid regime is based on pure thermal convection and excludes the dynamical effects of crust production that involves production and eruption of melt \citep[e.g.,][]{kite,noack2012coupling}. Crust production is rarely modelled in global mantle convection simulations since it is computationally more expensive \citep[see][for implementations]{xie04,keller09,nakagawa2010influence3D}, although \citet{moore13,lourenco16} reported that melting and (basaltic) crust production can have a first order impact on the convection regime of Earth-like planets. For example, strong enough eruptive magmatism can turn stagnant-lid regime into an episodic regime \citep{lourenco16}. 

Yet, melting and crust production will not always affect the convection regime. Here, we showed that partial melting can hardly occur on very large exoplanets  {that are in a stagnant-lid regime}. Although magmatism might be important for Earth-sized planets, it could be negligible for smaller mass (Mars-sized) or larger mass planets (super-Earths). Small planets cool much faster which makes melting only important in early stages as it has been shown in the case of Mars \citep{taylor2009planetary}.  {For high planet masses there are two effects that lead to reduced depletion. The first effect is the decrease in density cross-over depth with mass (see details in Section \ref{crossy}). The second effect is the increase of melting temperature with pressure. Thus, for high mass planets, the melting temperature beneath the lithosphere is generally higher than the adiabatic temperature which prevents melting.}
In such cases, magmatism might be restricted to planets with very thin lithosphere thicknesses that can develop in regimes such as plate tectonics. In the future,  {further investigations are necessary to better understand} the effect of different tectonic regimes on outgassing.

\section{Conclusions}
\label{Conclusions}
The atmospheres of the terrestrial Solar System planets are shaped by volcanic outgassing that occur on geological timescales, which we also expect to be relevant for super-Earth atmospheres. Furthermore, the atmospheres are the only parts of exoplanets that can be directly probed and upcoming missions (e.g., JWST, E-ELT) will provide detailed insights on exoplanet atmospheres. The interpretation of super-Earth atmospheres crucially relies on our understanding of volcanic outgassing. Here, we have thoroughly studied the diversity of  outgassing on stagnant-lid super-Earths given the anticipated diversity of their interiors. Thus, our study informs upcoming findings of observed super-Earth atmospheres.

Specifically, we investigated the amount of outgassed CO$_2$ given the anticipated diversity of super-Earths interiors.  We built on the work of \citet{noack17melting} and assumed a stagnant-lid convection regime. We accounted for a broad range of possible interiors of rocky exoplanets (1--8 \ME) that are in agreement with commonly observed astrophysical constraints of mass, radius, and stellar abundances. Stellar abundances of refractory elements are candidates for placing constraints on the relative abundance of rock-forming elements (i.e., Mg, Si , Fe) in the planet bulk. We also accounted for possible variations in interior parameters that are very difficult if not impossible to constrain from astrophysical data. These mostly include  {initial and} thermal parameters, e.g., the amount of radiogenic heat sources, the initial mantle temperature, or additional heat flux from the core;  {other investigated parameters are composition-related effects such as viscosity, influence of water and density-cross over pressure}. The surface temperatures were assumed to be Earth-like.

Our results are comparable to \citet{noack17melting}, where a simple silicate mantle and pure iron core composition was used, i.e., at high planetary masses outgassing ceases.

Based on our large number of  {2340} super-Earth models, we conclude the following:
\begin{itemize}
    \item Planetary mass $M_{\rm p}$ mainly influences the amount of outgassing on stagnant-lid planets. At small masses (< 2\ME, for the reference case), maximum mantle depletion is reached and outgassing positively correlates with planet mass, since it is controlled by the absolute mantle volume. At large masses (> 2\ME, for the reference case), depletion and thus outgassing decreases with planet mass, which is due to the increasing pressure gradient that  {leads to an increasing melting temperature beneath the lithosphere and} limits melting to shallower depths. 
    For stagnant-lid planets above $\sim$7 \ME, the large pressure gradient  {and the high melting temperatures beneath the lithosphere} generally prohibits partial melting at depth. Thus, for stagnant-lid planets, we expect that (1) there is a mass range of planets for which  {outgassing is most effective} 
    and (2) there is an upper mass limit above which outgassing rarely occurs. \rev{This predicted trend of CO$_2$ atmospheres with planet mass can be observationally tested for exoplanets.} Deviating behaviours  {may be explained by} dynamic regimes other than stagnant-lid, e.g., plate tectonics,  {or atmospheres being dominated by the early outgassing during the cooling of a magma ocean}.  The distribution of enriched atmospheres can be observationally tested with upcoming missions that aim at characterizing exoplanet atmospheres (e.g., JWST, E-ELT).
    \item Thermal parameters can significantly shift the mass range where maximum outgassing can occur and thus shift the transition between positive and negative correlation between $M_{\rm p}$  and $p_{\rm CO_2}$. We find that by varying the initial mantle temperature from 1600 K to 2000 K, this shift is on the order of 1 \ME, whereas the variation from 0.5 to 1.5 time the amounts of Earth-like radiogenic heat sources results in a shift on the order of 3 \ME. The tested ranges of thermal parameters broadly covers the expected variability among stagnant-lid exoplanets. 
    \item \rev{The anticipated range of exoplanet ages is wide and on the scale of Gyrs. Although, most of our results summarize the outgassing after 4.5 Gyrs of evolution, we discuss the evolution of volcanism up to 10 Gyrs (in Section \ref{Atimedep}). For ages below 4.5 Gyrs, planetary age can have first-order influence on depletion and the amount of outgassed CO$_2$. However, outgassing beyond 4.5 Gyrs does only add small or negligible amounts of CO$_2$ to an atmosphere. Our investigation shows that planets of masses above 3 \ME (reference case) do not have significant outgassing, even over an extended evolution of 10 Gyrs. 
    }
    \item Mantle composition seems to be of secondary influence for outgassing. Mantle composition influences melting temperature and mantle density.  A more iron-rich mantle material has a lower melting temperature which increases melting and thus leads to higher outgassing ($<$ 2\ME, for the reference case). At the same time, an iron-rich mantle composition implies high mantle densities which increases the pressure gradient in the lithosphere. Thus at large masses ($>$ 2\ME, for the reference case) when melting is limited by the pressure gradient as discussed earlier, a more iron-rich mantle tends to outgas less.  {Composition also influences the viscosity of the mantle, the melting temperature and the density-cross over pressure. While all of these factors tend to influence the amount of outgassing in the intermediate mass range (2-4\ME), no significant change in outgassing is observed for low-mass planets, where depletion is efficient for all tested cases as well as for more massive planets, where little or no outgassing occurs.}
    \item The effect  of core size is of secondary influence for outgassing. At small masses, where outgassing is controlled by mantle volume, a smaller core size increases the amount of outgassing. At larger masses, we find the opposite trend. Due to the bulk abundance constraints, a smaller core implies an iron-rich and thus dense mantle material, which results in a higher pressure gradient in the lithosphere. Therefore, melting is limited to shallower regions and outgassing is reduced. 
    \item We estimate the respective gas layer thicknesses of the calculated outgassed CO$_2$ and compare them with independent estimates of Venus and Mars and find good agreement.
\end{itemize}

Finally, we provide scaling laws that summarize the influence of first- and second-order interior parameters on mantle depletion and outgassing on stagnant-lid planets. 
 Thereby, our study represents a significant step towards providing interpretative means for comparative studies of exoplanet atmospheres.

\begin{acknowledgements}
C.D. received funding from the Swiss National Foundation under grant 200020\_160120, PZ00P2\_174028, and from the MERAC grant by the Swiss Society of Astrophysics and Astronomy. L.N. received funding from the Interuniversity Attraction Poles Programme initiated by the Belgian Science Policy Office through the Planet Topers alliance. 
A.B.R. received funding from the European Research Council under the European Union's Seventh Framework Programme (FP/20072013)/ERC Grant Agreement number 320639 project iGEO. This study was in part carried out within the frame of the National Centre for Competence in Research PlanetS. 
\end{acknowledgements}

\bibliographystyle{aa}
\bibliography{DornBib} 

\end{document}